\documentclass[aps,pre,twocolumn]{revtex4-1}
\usepackage[dvips]{graphicx}
\usepackage{amsmath}
\usepackage{verbatim}
\newcommand{\la}{\left\langle}
\newcommand{\ra}{\right\rangle}

\newcommand{\etalia}{{\it et al.~}}

\newcommand{\PRL}{Phys.~Rev.~Lett.~}

\begin{document}

\title{Osmotic Pressure of Permeable Ionic Microgels: Poisson-Boltzmann Theory 
and Exact Statistical Mechanical Relations in the Cell Model}

\author{Alan R. Denton}
\email[]{alan.denton@ndsu.edu}
\author{Mohammed O. Alziyadi}
\affiliation{Department of Physics, North Dakota State University,
Fargo, ND 58108-6050, USA}

\vspace*{0.6cm}

\begin{abstract}
Ionic microgels are soft colloidal particles, composed of crosslinked polymer networks,
that ionize and swell when dispersed in a good solvent. Swelling of these permeable,
compressible particles involves a balance of electrostatic, elastic, and mixing contributions
to the single-particle osmotic pressure. The electrostatic contribution depends on the
distributions of mobile counterions and coions and of fixed charge on the polymers.
Within the cell model, we employ two complementary methods to derive the electrostatic
osmotic pressure of ionic microgels. In Poisson-Boltzmann (PB) theory, we minimize a
free energy functional with respect to the electrostatic potential to obtain the bulk pressure.
From the pressure tensor, we extract the electrostatic and gel contributions to the
total pressure. In a statistical mechanical approach, we vary the free energy with respect
to microgel size to obtain exact relations for the microgel electrostatic osmotic pressure.
We present results for planar, cylindrical, and spherical geometries. For models of 
membranes and microgels with fixed charge uniformly distributed over their surface or volume, 
we derive analogues of the contact value theorem for charged colloids. We validate these 
relations by solving the PB equation and computing ion densities and osmotic pressures. 
When implemented within PB theory, the two methods yield identical electrostatic osmotic 
pressures for surface-charged microgels. For volume-charged microgels, the exact 
electrostatic osmotic pressure equals the average of the corresponding PB profile over 
the gel volume. We demonstrate that swelling of ionic microgels depends on variation 
of the electrostatic pressure inside the particle and discuss implications for 
interpreting experiments.
\end{abstract}

\maketitle
\newpage



\section{Introduction}

For over a century, Poisson-Boltzmann (PB) theory has been widely applied to model interfaces 
between charged surfaces and ionic solutions~\cite{Hunter,Israelachvili1992,EvansWennerstrom}. 
By accounting for thermal motion of ions, the theory explains the origin of a 
diffuse electrical double layer comprising fixed surface charges and mobile microions
(counterions, coions) in solution. The densities of microions are described by the PB equation, 
which results from combining the Poisson equation for the electrostatic potential with 
Boltzmann distributions for the microion charge densities.
Despite neglecting interparticle correlations through its underlying mean-field approximation,
PB theory often reasonably describes the distributions of weakly correlated (usually monovalent) 
microions in solution around macroions of various types, including charged colloids, 
polyelectrolytes, and charged lipid bilayers.
 
Combining the PB theory of electrostatic interactions between charged surfaces with the 
Hamaker theory of van der Waals interactions yields the famous Derjaguin-Landau-Vervey-Oberbeek 
(DLVO) theory~\cite{DL1941,VO1948}, which explains the stability of charge-stabilized 
colloidal suspensions~\cite{Hunter,Israelachvili1992,EvansWennerstrom}. 
PB theory also provides a basis for modern theories of effective electrostatic interactions 
in soft matter~\cite{likos2001,denton-zvelindovsky2007,denton-cecam2014,denton2010,denton2003}.
Following important elaborations of the original Gouy-Chapman formulation~\cite{Gouy1910,Chapman1913} 
of PB theory by Stern~\cite{Stern1924} and Grahame~\cite{Grahame1947} to account for 
the nonzero size, hydration, and surface adsorption of ions, recent refinements of the theory 
have addressed ion-specific effects by incorporating nonuniform dielectric permittivity
and non-Coulombic interactions~\cite{Ben-Yaakov-Andelman-Harries-Podgornik-jpcm2009,
Ben-Yaakov-Andelman-Harries-Podgornik-jpcb2009,Markovich-Andelman-Podgornik2016,
MayPRE2012,
MayJCP2013,
MayLangmuir2015,
MayACIS2017,
May2018}.

In practical applications, PB theory is often implemented in the 
cell model~\cite{marcus1955,wennerstrom1982,deserno-holm2001}, 
which abstracts from a bulk suspension a single macroion and its associated microions. 
The shape of the macroion dictates the appropriate geometry of the cell (planar, cylindrical,
spherical), the symmetry of which eases the computational solution of the nonlinear PB equation.
Predictions for microion densities and osmotic pressure prove to be most accurate for systems 
with relatively low salt concentration~\cite{denton2010,hallez2014}.

Aside from determining electrical properties of charged surfaces in ionic solutions, 
such as differential capacitance and electrostatic interactions between surfaces, 
microion densities have relevance for bulk thermodynamic properties of charge-stabilized 
colloidal suspensions, such as the pressure and phase behavior. In fact, within the cell model, 
the pressure is proportional to the total microion density at the edge of the cell.
This important relation was first established in the framework of PB theory~\cite{marcus1955}, 
but was subsequently proven to be exact in the cell model~\cite{wennerstrom1982}, 
independent of PB theory. 

In recent years, PB theory has been widely applied to penetrable macroions, which are 
distinguished by their permeability to solvent and microions. The most intensively studied 
are microgels -- soft colloidal particles consisting of crosslinked polymer networks 
-- whose ability to swell and deswell in solution leads to unusual materials properties
with a multitude of practical applications, including drug delivery, water filtration, and
biosensing~\cite{MicrogelBook2011,HydrogelBook2012,lyon-nieves-AnnuRevPhysChem2012,karg-richtering2019}.
Predictions of PB theory for the distribution of monovalent counterions in the cell model 
of ionic microgels often prove accurate when compared with molecular simulations~\cite{holm2009,
molina2013,hedrick-chung-denton2015,denton-tang2016,holm2019}.
Relatively little attention has been devoted, however, to modeling thermodynamic properties, 
such as osmotic pressure~\cite{levin2002,gottwald2005,likos2011,colla-likos2014}. 
The electrostatic contribution to the interior osmotic pressure of a permeable macroion, 
defined as the difference in pressure between the inside and outside, is determined by the 
density profiles of the fixed charge and mobile microions. The interior osmotic pressure, 
in turn, determines the equilibrium swollen size of an ionic microgel through a balance 
between electrostatic, elastic, and mixing contributions to the total osmotic 
pressure~\cite{cloitre-leibler1999,cloitre-leibler2003,vlassopoulos-cloitre2014}.
Accurately predicting the structure and phase behavior of concentrated microgel suspensions 
requires consistently incorporating deswelling of the compressible 
particles~\cite{urich-denton2016,weyer-denton2018}.

Swelling of ionic microgels has been probed in recent years by a variety of experimental methods, 
including light scattering~\cite{nieves-macromol2000,nieves-jcp2001,nieves-jcp2003,weitz-jcp2012,
schurtenberger-ZPC2012,holmqvist-shurtenberger2012,braibanti-perez2016,Nojd2018},
neutron scattering~\cite{Nojd2018,nieves-jcp2010},
electrophoresis~\cite{nieves-jcp2005},
and osmometry~\cite{nieves-prl2015}.
Measurements of the swollen radii of pNIPAM microgels in deionized 
solutions~\cite{holmqvist-shurtenberger2012,braibanti-perez2016,Nojd2018}
indicate that deswelling sets in at concentrations well below random close packing, 
where particles start to overlap, qualitatively consistent with theoretical predictions 
for coarse-grained models~\cite{denton-tang2016,weyer-denton2018}.
Such observations point to the vital role of electrostatic interactions in driving 
deswelling of ionic microgels.

While theories of macroscopic ionic gels are well established~\cite{katchalsky1951,katchalsky1955,
barrat-joanny-pincus1992,khokhlov1993,khokhlov1997,rubinstein-dobrynin1996}, a correspondingly 
comprehensive description of ionic microgels is still lacking. In particular, the important distinction 
between the osmotic pressure of a bulk suspension and the osmotic pressure of a single microgel 
is neither widely appreciated nor fully understood. As a result, simplified approximations 
are often adopted for the electrostatic component of the interior osmotic pressure 
in lieu of a more rigorous theory capable of quantitative predictions. 
The purpose of this paper is to fill this lacuna by deriving general expressions for the 
electrostatic component of the osmotic pressure of permeable macroions and to discuss implications 
for swelling and thermodynamic phase behavior of microgels.

The remainder of the paper is organized as follows.
In Sec.~\ref{models}, we define the primitive and cell models of permeable macroions.
In Sec.~\ref{theory}, we derive from Poisson-Boltzmann theory and the pressure tensor 
general expressions for the bulk pressure and the electrostatic component of the 
osmotic pressure of ionic microgels in the planar, cylindrical, and spherical cell models.
In Sec.~\ref{applications}, we apply PB theory to derive explicit expressions for the 
electrostatic component of the osmotic pressure of surface- and volume-charged microgels. 
In Sec.~\ref{exact}, we derive exact statistical mechanical relations in the cell model 
for the electrostatic osmotic pressure of ionic microgels. 
In Sec.~\ref{validation}, we validate our results by numerically solving the PB equation to compute 
pressures and comparing predictions of PB theory with the corresponding exact relations. 
We also discuss the relevance of our results for modeling swelling of ionic microgels and
interpreting experiments.
In Sec.~\ref{conclusions}, we summarize and conclude with an outlook for future work.

\section{Models}\label{models}

\subsection{Primitive Model}\label{primitive-model}

We consider an aqueous suspension of $N_m$ ionic microgels in a volume $V$ at temperature $T$. 
Each particle is modeled as a swollen slab, cylinder, or sphere, composed of a crosslinked 
network of polymer chains. Dissociation of $Z$ counterions from ionizable sites on the 
polymers leaves each microgel with a charge $-Ze$, where $e$ is the electron charge. 
The spatial distribution of fixed charge on the polymers depends on the distribution of 
monomers. The presence of salt in solution contributes coions and additional counterions.  
In chemical (Donnan) equilibrium with an electrolyte reservoir of salt ion pair density $n_0$, 
the suspension contains $N_s$ dissociated salt ion pairs. For simplicity, we assume the microions 
are monovalent and pointlike. From the condition of electroneutrality, the total number of 
counterions is $N_+=ZN_m+N_s$ and coions $N_-=N_s$. 

The microgels are permeable to both microions and solvent (water). In the primitive model, 
the solvent is a dielectric continuum of uniform dielectric constant $\epsilon$, which reduces 
the strength of the bare Coulomb interaction between a pair of ions at separation $r$ to 
$v(r)=e^2/(\epsilon r)$ (Gaussian units).
The neglect of solvent structure limits the primitive model to phenomena that 
do not depend significantly on ion hydration effects.

\vspace*{-0.2cm}
\subsection{Cell Model}\label{cell-model}

\begin{figure}
\includegraphics[width=0.9\columnwidth]{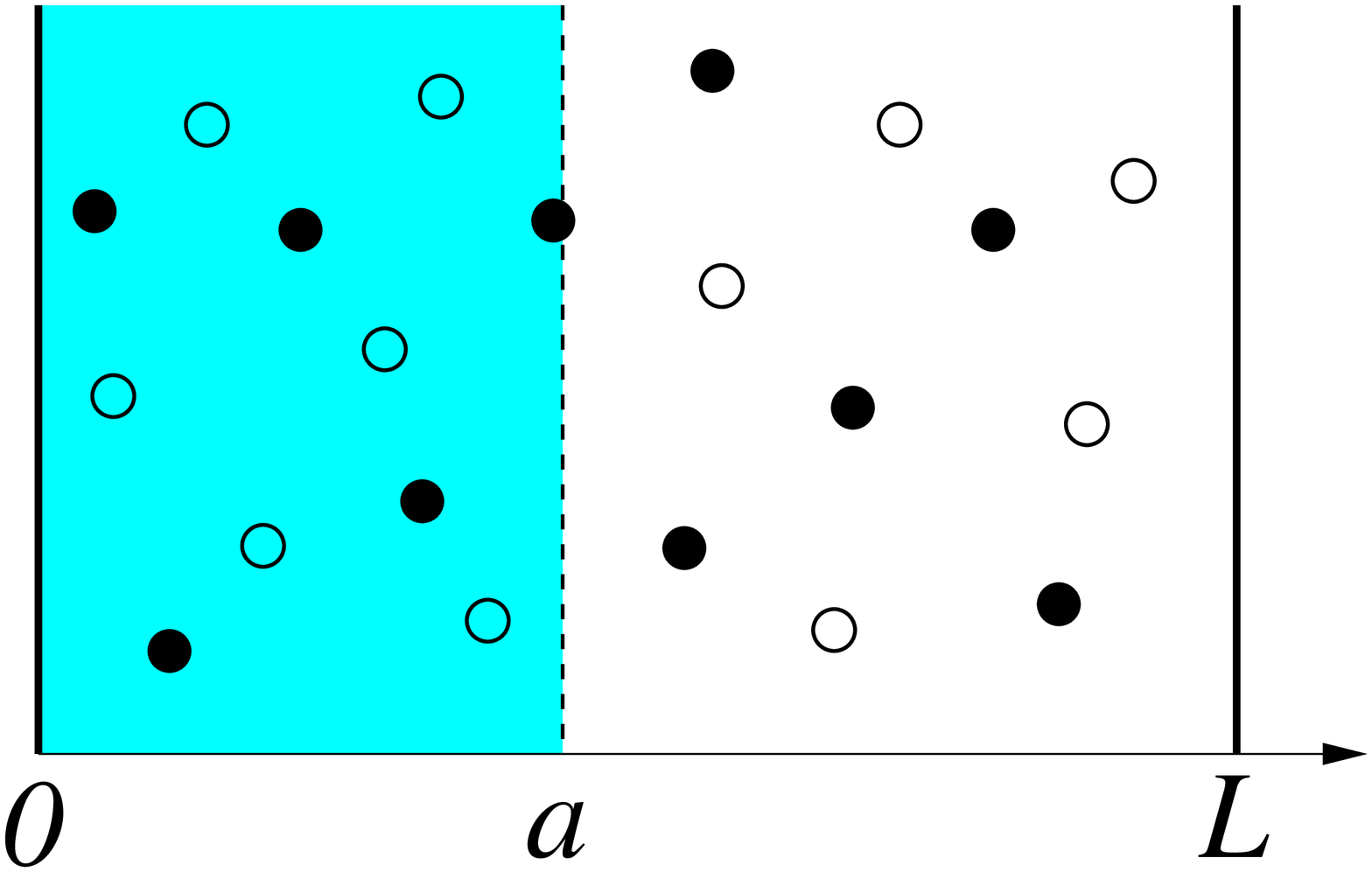} \\[2ex]
\includegraphics[width=0.55\columnwidth]{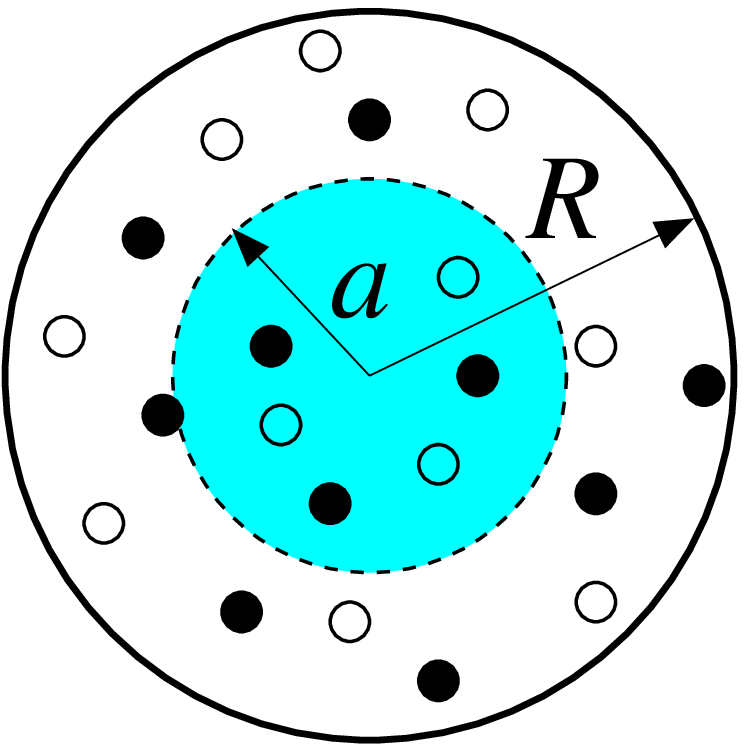}
\vspace*{-0.2cm}
\caption{
Schematic drawings of permeable ionic microgels in the cell model. Planar microgel (blue slab) 
of swollen thickness $a$ and cylindrical or spherical microgel (blue disk) of swollen radius $a$, 
oppositely-charged counterions (filled black disks), like-charged coions (unfilled disks), 
and implicit solvent (dielectric continuum) in a cell of width $L$ or radius $R$. 
}\label{cell-model-fig}
\end{figure}

The cell model~\cite{marcus1955,wennerstrom1982,deserno-holm2001} represents a bulk suspension 
of macroions, by a single macroion, along with mobile microions, in a cell of commensurate symmetry
(Fig.~\ref{cell-model-fig}). In the planar cell model, a slab of ionic gel of thickness $a$ 
(infinite in the other two dimensions)
with fixed charge distribution $-en_f(x)$ is placed at one side of a cell of width $L$, representing 
a suspension of planar microgels of thickness $2a$ and volume fraction $\phi=a/L$. 
In the cylindrical cell model, an infinitely long cylinder of ionic gel of cross-sectional radius $a$ 
with axially symmetric fixed charge distribution $-en_f(r)$ is centered in a cylindrical cell 
of radius $R$, representing a suspension of cylindrical microgels~\cite{Raghavan2018}
of volume fraction $\phi=(a/R)^2$. 
In the spherical cell model, a sphere of ionic microgel of radius $a$ with spherically symmetric 
fixed charge distribution $-en_f(r)$ is centered in a spherical cell of radius $R$, representing 
a suspension of spherical microgels of volume fraction $\phi=(a/R)^3$. 

The counterions and coions have equilibrium number density profiles $n_+(r)$ and $n_-(r)$,
respectively, and corresponding charge density profiles $\pm en_{\pm}(r)$,
which are determined by the fixed charge distribution. Electroneutrality
of the system dictates that the electric field vanishes at the cell boundaries: $x=0$ and $L$
for the planar cell and $r=R$ for the cylindrical and spherical cells. Combining the cell model 
with the primitive model, the solvent is represented by a dielectric continuum. 
While a significant abstraction from a bulk suspension of macroions, the cell model has proven 
accurate, compared with multi-macroion simulations and experiments, in predicting osmotic pressures 
of deionized suspensions of charged colloids~\cite{denton-cecam2014,denton2010} and
ionic microgels~\cite{hedrick-chung-denton2015}. Suspensions of charged clay platelets 
have been successfully modeled using the Wigner-Seitz cell model~\cite{trizac-hansen1997}, 
with circular or square platelets placed at the center of a cylindrical or parallelepipedic 
cell. In the latter geometry, the fixed charge distribution, electric field, and 
microion distributions are not axially symmetric.

\section{Poisson-Boltzmann Theory}\label{theory}

\subsection{Free Energy and Osmotic Pressure}

For a suspension of fixed volume and temperature in Donnan equilibrium with a salt reservoir, 
imposing a salt ion chemical potential of $\mu_0=k_BT\ln n_0$, the appropriate thermodynamic potential 
is the semi-grand potential $\Omega$, a Legendre transform of the Helmholtz free energy $F$:
\begin{equation}
\Omega[n_{\pm}]=F[n_{\pm}]-\mu_0\int_V dV\, (n_+ + n_-).
\label{Omega-def}
\end{equation}
In a mean-field approximation, the electrostatic part of the semi-grand potential functional 
can be expressed as
\begin{equation}
\Omega[n_{\pm}]=\Omega_{\rm id}[n_{\pm}]+\frac{e}{2}\int_V dV\, (n_+-n_--n_f)\varphi,
\label{Omega}
\end{equation}
where the exact ideal-gas functional (in $k_BT$ units) is
\begin{equation}
\Omega_{\rm id}=
\int_V dV\left\{n_+\left[\ln\left(\frac{n_+}{n_0}\right)-1\right] 
+n_-\left[\ln\left(\frac{n_-}{n_0}\right)-1\right]\right\}
\label{Omegaid}
\end{equation}
and the electrostatic potential $\varphi$ is related to all of the ion densities via
\begin{equation}
\varphi({\bf r})=\frac{1}{2}\int_V d{\bf r}'\,[n_+({\bf r}')-n_-({\bf r}')-n_f({\bf r}')]
\frac{e}{|{\bf r}-{\bf r}'|},
\label{phi}
\end{equation}
with the integrals extending over the system volume $V$.
To simplify notation, we henceforth express all energies in thermal ($k_BT$) units.
For given distributions of fixed and mobile charges, $\varphi$ obeys the Poisson equation: 
\begin{equation}
\nabla^2\varphi=-\frac{4\pi e}{\epsilon}(n_+-n_--n_f),
\label{Poisson1}
\end{equation}
or, upon introducing the reduced electrostatic potential,
$\psi\equiv e\varphi/(k_BT)$, and Bjerrum length, $\lambda_B=e^2/(\epsilon k_BT)$, 
\begin{equation}
\nabla^2\psi=-4\pi\lambda_B(n_+-n_--n_f).
\label{Poisson2}
\end{equation}
Minimizing $\Omega$ with respect to the microion densities, i.e., requiring 
$\delta \Omega/\delta n_{\pm}=0$, generates the Boltzmann approximations, 
$n_{\pm}=n_0\exp(\mp\psi)$, which when substituted into 
Eq.~(\ref{Poisson2}) yield the PB equation~\cite{lowen92,lowen_jcp93},
\begin{equation}
\nabla^2\psi=\kappa^2\sinh\psi+4\pi\lambda_B n_f,
\label{PB}
\end{equation}
where $\kappa=\sqrt{8\pi n_0\lambda_B}$ is the Debye screening constant. 
Combining Eqs.~(\ref{Omega}), (\ref{Omegaid}), and (\ref{PB}), we have 
\begin{eqnarray}
\Omega&=&\int_V dV\, \left(2n_0\left(\psi\sinh\psi-\cosh\psi\right)
-\frac{\psi\nabla^2\psi}{8\pi\lambda_B}\right)
\nonumber\\[1ex]
&=&\int_S dS\, \frac{\psi\nabla\psi}{8\pi\lambda_B}
-\int_V dV \left(\frac{|\nabla\psi|^2}{8\pi\lambda_B}
+2n_0\cosh\psi+n_f\psi\right).
\nonumber\\
\label{OmegaPB2}
\end{eqnarray}
As the electric field (-$\nabla\psi$) vanishes on the boundary $S$,
\begin{equation}
\Omega=-\int_V dV\, \left(\frac{|\nabla\psi|^2}{8\pi\lambda_B}+2n_0\cosh\psi+n_f\psi\right).
\label{OmegaPB3}
\end{equation}
In applications to hard charged colloids, the fixed charge is usually confined to a surface,
while for ionic gels, $n_f$ may be a volume-distributed charge. Legendre transforming to the 
Gibbs free energy, $G=\Omega+PV$, shifts the constraint from fixed volume $V$ to fixed 
thermodynamic pressure $P$ (in $k_BT$ units), which may be determined by variational minimization 
of $G$ with respect to $\psi$~\cite{Ben-Yaakov-Andelman-Harries-Podgornik-jpcb2009,
Ben-Yaakov-Andelman-Harries-Podgornik-jpcm2009,Markovich-Andelman-Podgornik2016}.
In cell models with planar, cylindrical, or spherical symmetry, $P$ is the normal force 
per unit area acting on the outer boundary of the cell -- a surface whose normal is parallel to 
the electric field~\cite{Widom2009} -- corresponding to the bulk pressure of the suspension. 

It is important to distinguish the bulk pressure from the pressure tensor in the cell, 
the elements of which represent forces per unit area exerted on surfaces whose normal vectors 
are oriented parallel to different coordinate axes. In a dielectric medium with electric field 
${\bf E}$ and total number density of mobile ions $n(r)$, the electrostatic pressure tensor 
(in $k_BT$ units) is given by~\cite{landau-lifshitz1984,ljunggren-eriksson1988,
trizac-hansen1997,luo2013,luo2014,holm2019}
\begin{equation}
{\bf P}_e=\frac{\epsilon}{4\pi k_BT}\left(\frac{1}{2}E^2{\bf 1}
-{\bf E}\otimes{\bf E}\right)+n(r){\bf 1},
\label{Pe-tensor}
\end{equation}
where ${\bf 1}$ is the unit tensor.
The first term on the right side is the Maxwell pressure tensor (negative of the 
Maxwell stress tensor), associated with the electric field generated by both the 
fixed charge and the mobile microions. The second term is the kinetic pressure tensor,
associated with the momentum transported by the mobile microions. 
The normal electrostatic pressure, acting on a surface whose normal is parallel to ${\bf E}$, is
\begin{equation}
P_{e,\parallel}(r)=-\frac{|\psi'(r)|^2}{8\pi\lambda_B}+n(r),
\label{Pe-normal}
\end{equation}
while the electrostatic pressure in transverse directions is
\begin{equation}
P_{e,\bot}(r)=\frac{|\psi'(r)|^2}{8\pi\lambda_B}+n(r).
\label{Pe-transverse}
\end{equation}
In a gel medium, the {\it total} pressure tensor is the sum of the electrostatic and
gel pressure tensors, ${\bf P}={\bf P}_e+{\bf P}_g$.
The bulk pressure equals the normal component of the total pressure at the cell edge:
$P=P_{\parallel}(R)={\bf P}_{rr}(R)$.

In matrix notation, the total pressure tensor can be expressed as 
\begin{equation}
\left({\bf P}\right)=
\left(
\begin{array}{ccc} 
P_{\parallel} & 0 & 0 \\
0 & P_{\bot} & 0 \\
0 & 0 & P_{\bot}
\end{array}
\right)
\label{P-matrix}
\end{equation}
where we choose the first coordinate parallel to ${\bf E}$ and
off-diagonal elements are zero in the absence of shear forces.
The condition of mechanical equilibrium is that the net force density at any point 
must vanish, which is equivalent to vanishing of the divergence of the total pressure tensor:
$\nabla\cdot{\bf P}=0$. 
As shown below, this condition implies that the elements of the total pressure tensor, 
$P_{\parallel}$ and $P_{\bot}$, are spatially constant in the planar cell model, 
but spatially varying in cylindrical and spherical cells.

It is also essential here to distinguish between the terms ``pressure" and ``osmotic pressure," 
which are often used interchangeably in soft matter physics. Let us avoid confusion by defining 
the osmotic pressure of a suspension as the difference in bulk pressure between 
the suspension and an electrolyte reservoir (the pressure itself being measured relative 
to vacuum). This bulk property is distinct from the osmotic pressure of a single microgel,
defined as the change in the normal component of the total pressure tensor across
the surface of the gel. In equilibrium, the polymer network swells until the 
single-microgel osmotic pressure vanishes, implying continuity of the normal component 
of the total pressure at the microgel surface. Next, we derive a number of 
fundamental expressions for the osmotic pressure in cell models of various geometries.

\subsection{Planar Cell Model}\label{planar-cell-model}

In a planar cell of width $L$, we can express the Gibbs free energy functional per unit area
in the form
\begin{equation}
G=-\int_0^L dx\, g(x,\psi,\psi'),
\label{G-plane}
\end{equation}
where
\begin{equation}
g(x,\psi,\psi')=\frac{|\psi'(x)|^2}{8\pi\lambda_B}+2n_0\cosh\psi(x)+n_f(x)\psi(x)-P(L),
\label{g-plane}
\end{equation}
depending explicitly on $x$ through the fixed charge $n_f(x)$.
To emphasize the distinction between the bulk pressure and the total pressure inside the cell,
we denote the former by $P(L)$ and the latter by $P(x)$, although, as it turns out
in planar geometry, $P(x)=P(L)$.
Minimizing the functional with respect to $\psi$, i.e., 
replacing $\psi(x)$ with $\psi(x)+\alpha\eta(x)$, with arbitrary perturbation function $\eta(x)$,
and requiring $dG/d\alpha=0$, leads to~\cite{RHB} 
\begin{eqnarray}
\delta G&=&\int_0^L dx\, \eta
\left[\frac{\partial g}{\partial\psi}-\frac{d}{dx}\frac{\partial g}{\partial\psi'}\right]
+\left[\eta\frac{\partial g}{\partial\psi'}\right]_0^L
\nonumber\\[1ex]
&+&\left[g-\psi'\frac{\partial g}{\partial\psi'}\right]_{x=L}\Delta x=0,
\label{deltaG}
\end{eqnarray}
where $\Delta x$ represents a variation of the endpoint ($x=L$), which allows the cell 
to adjust its volume to reach equilibrium at the externally applied normal pressure.
Equation~(\ref{deltaG}) immediately yields the Euler-Lagrange equation
\begin{equation}
\frac{\partial g}{\partial\psi}-\frac{d}{dx}\frac{\partial g}{\partial\psi'}=0,
\label{ELx}
\end{equation}
the boundary conditions
\begin{equation}
\frac{\partial g}{\partial\psi'}\Big |_{x=0}=
\frac{\partial g}{\partial\psi'}\Big |_{x=L}=0,
\label{dg-dpsip}
\end{equation}
and the constraint
\begin{equation}
\left[g-\psi'\frac{\partial g}{\partial\psi'}\right]_{x=L}=0.
\label{dg-dx1}
\end{equation}
Substituting $g$ from Eq.~(\ref{g-plane}) into Eqs.~(\ref{ELx})-(\ref{dg-dx1}) 
recovers the PB equation [Eq.~(\ref{PB})] and 
leads to $\psi'(0)=\psi'(L)=0$ (implying electroneutrality) and also the 
thermodynamic pressure acting at the edge of the cell:
\begin{equation}
P(L)=-\frac{|\psi'(L)|^2}{8\pi\lambda_B}+2n_0\cosh\psi(L)+n_f(L)\psi(L).
\label{P-plane0}
\end{equation}
Assuming $n_f(L)=0$ (no fixed charge at the cell edge), we finally obtain
\begin{equation}
P(L)=2n_0\cosh\psi(L)=n_+(L)+n_-(L),
\label{cell-theorem-plane}
\end{equation}
which is the well-known planar cell theorem relating the bulk pressure of the suspension
to the ion density at the cell edge~\cite{marcus1955,wennerstrom1982}. 
The osmotic pressure of the suspension is obtained by subtracting the pressure of the reservoir. 

A more general expression for the bulk pressure, applicable to any fixed charge distribution, 
follows by first using Eq.~(\ref{ELx}) to simplify the total derivative of $g$:
\begin{equation}
\frac{dg}{dx}=\frac{\partial g}{\partial x}+\psi'\frac{\partial g}{\partial\psi}
+\psi''\frac{\partial g}{\partial\psi'}
=\frac{\partial g}{\partial x}+\frac{d}{dx}\left(\psi'\frac{\partial g}{\partial\psi'}\right).
\label{dg-dx2}
\end{equation}
Integrating from position $x$ to the cell edge ($x=L$) yields
\begin{equation}
\left(g-\psi'\frac{\partial g}{\partial\psi'}\right)\Big |_x^L
=\int_x^L du\, \frac{\partial g}{\partial u}.
\label{dg-dx3}
\end{equation}
Substituting $g$ from Eq.~(\ref{g-plane}) into Eq.~(\ref{dg-dx3}), noting that
$\partial g/\partial u=n_f'(u)\psi(u)$, integrating by parts, and making use of 
Eqs.~(\ref{P-plane0}) and (\ref{cell-theorem-plane}), we obtain 
\begin{equation}
P(L)=-\frac{|\psi'(x)|^2}{8\pi\lambda_B}+2n_0\cosh\psi(x)+P_g(x),
\label{P-plane2}
\end{equation}
where the first two terms on the right side are the Maxwell pressure 
and the kinetic pressure of the mobile microions and the third term,
\begin{equation}
P_g(x)=-\int_x^L du\, n_f(u)\psi'(u),
\label{Pg-plane}
\end{equation}
is the counteracting gel pressure required to stabilize the fixed charge against 
the electric field ($-\psi'$), thus ensuring mechanical stability of the gel.
For charged surfaces, the third term in Eq.~(\ref{P-plane2}) is often implicit,
subsumed into the boundary conditions~\cite{Markovich-Andelman-Podgornik2016}, 
while for ionic gels with volume-distributed fixed charge it must be explicitly included.
Note that Eq.~(\ref{Pg-plane}) is consistent with the force-balance condition for 
mechanical equilibrium of the fixed charge in the presence of an electric field:
\begin{equation}
P_g'(x)=n_f(x)\psi'(x).
\label{dPg-dx}
\end{equation}
In microgels, the gel pressure would be provided by the swollen polymer network,
while for a membrane, $P_g$ could be supplied by the restraining force 
of an external support or a lipid bilayer. 
The normal component of the electrostatic pressure is then identified as
\begin{equation}
P_{e,\parallel}(x)=-\frac{|\psi'(x)|^2}{8\pi\lambda_B}+2n_0\cosh\psi(x),
\label{Pe-plane1}
\end{equation}
consistent with Eq.~(\ref{Pe-normal}) in which 
$n(x)=2n_0\cosh\psi(x)$ is the total microion number density.
From Eqs.~(\ref{cell-theorem-plane}) and (\ref{Pg-plane}), we can also write
\begin{equation}
P_{e,\parallel}(x)=n_+(L)+n_-(L)+\int_x^L du\, n_f(u)\psi'(u).
\label{Pe-plane2}
\end{equation}
Note that in the planar cell, the condition for mechanical equilibrium,
$\nabla\cdot{\bf P}=0$, implies $dP(x)/dx=0$, and thus the total pressure 
inside the cell is spatially constant, i.e., $P(x)=P(L)$. In contrast, 
in cylindrical and spherical cell models, the total pressure inside the cell
is spatially varying, as shown in the next section.

\subsection{Cylindrical Cell Model}\label{cylindrical-cell-model}

A similar development yields the bulk pressure of a suspension modeled by
an infinitely long, axially symmetric, cylindrical cell of cross-sectional radius $R$.
The Gibbs free energy functional per unit length along the cylinder axis now takes the form
\begin{equation}
G=-2\pi\int_0^R dr\, g(r,\psi,\psi'),
\label{G-cylinder}
\end{equation}
where $r$ is the radial distance from the axis and 
\begin{eqnarray}
g(r,\psi,\psi')&=&r\left[\frac{|\psi'(r)|^2}{8\pi\lambda_B}+2n_0\cosh\psi(r)\right.
\nonumber\\[1ex]
&+&\left.\frac{}{}n_f(r)\psi(r)-P(R)\right].
\label{g-cylinder}
\end{eqnarray}
To again emphasize the distinction between the bulk pressure and the total pressure 
inside the cell, we denote the former by $P(R)$ and the latter by $P(r)$.
The additional factor of $r$ in Eq.~(\ref{g-cylinder}), from axial symmetry in cylindrical geometry, 
modifies the results derived in planar geometry.
Minimizing the functional with a variable endpoint ($r=R$)~\cite{RHB} 
yields the Euler-Lagrange equation
\begin{equation}
\frac{\partial g}{\partial\psi}-\frac{d}{dr}\frac{\partial g}{\partial\psi'}=0,
\label{ELr}
\end{equation}
the boundary conditions
\begin{equation}
\frac{\partial g}{\partial\psi'}\Big |_{r=0}=
\frac{\partial g}{\partial\psi'}\Big |_{r=R}=0,
\label{dg-dpsip-cylinder}
\end{equation}
and the constraint
\begin{equation}
\left(g-\psi'\frac{\partial g}{\partial\psi'}\right)\Big |_{r=R}=0.
\label{dg-dr1}
\end{equation}
Substituting $g$ from Eq.~(\ref{g-cylinder}) into Eqs.~(\ref{ELr})-(\ref{dg-dr1}) 
recovers the PB equation [Eq.~(\ref{PB})] in cylindrical polar coordinates,
the electroneutrality condition, $\psi'(R)=0$, and further yields the bulk pressure
(acting on the cell boundary)
\begin{equation}
P(R)=-\frac{|\psi'(R)|^2}{8\pi\lambda_B}+2n_0\cosh\psi(R)+n_f(R)\psi(R).
\label{P-cylinder0}
\end{equation}
In passing, we note that this result is consistent with the condition for
mechanical equilibrium of the cell~\cite{trizac-raimbault1999},
\begin{equation}
\frac{\partial P}{\partial\psi}=-(n_+-n_--n_f),
\label{dPdpsi}
\end{equation}
when the fixed charge of the ionic gel is included in the total charge density.
Assuming $n_f(R)=0$ (no fixed charge at the cell edge) leads to the 
cell theorem for the bulk pressure~\cite{marcus1955,wennerstrom1982}:
\begin{equation}
P(R)=2n_0\cosh\psi(R)=n_+(R)+n_-(R).
\label{cell-theorem-cylinder}
\end{equation}
As in planar geometry, we can derive a more general expression for the bulk pressure
by first using Eq.~(\ref{ELr}) to simplify the total derivative of $g$:
\begin{equation}
\frac{dg}{dr}=\frac{\partial g}{\partial r}+\psi'\frac{\partial g}{\partial\psi}
+\psi''\frac{\partial g}{\partial\psi'}
=\frac{\partial g}{\partial r}+\frac{d}{dr}\left(\psi'\frac{\partial g}{\partial\psi'}\right).
\label{dg-dr2}
\end{equation}
Integrating Eq.~(\ref{dg-dr2}) from radius $r$ to the cell edge ($r=R$) and using 
Eq.~(\ref{g-cylinder}) to evaluate $\partial g/\partial r$, we have 
\begin{equation}
\left(g-\psi'\frac{\partial g}{\partial\psi'}\right){\Big |}_r^R=\int_r^R du\, \left(\frac{g}{u}
+u n_f'(u)\psi(u)\right).
\label{P-cylinder1}
\end{equation}
Substituting $g$ from Eq.~(\ref{g-cylinder}) on both sides of Eq.~(\ref{P-cylinder1}), 
using Eq.~(\ref{P-cylinder0}), integrating by parts, and solving for $P(R)$, 
we obtain the following completely general expression 
for the bulk pressure in the cylindrical cell model: 
\begin{widetext}
\begin{equation}
P(R)=\frac{r}{R}\left(-\frac{|\psi'(r)|^2}{8\pi\lambda_B}+n(r)\right)
+\frac{1}{R}\int_r^R du\, \left(\frac{|\psi'(u)|^2}{8\pi\lambda_B}
+n(u)-u n_f(u)\psi'(u)\right),
\label{P-cylinder2a}
\end{equation}
\end{widetext}
where $n(r)\equiv n_+(r)+n_-(r)=2n_0\cosh\psi(r)$ is the total microion number density.
Setting $r=R$ in Eq.~(\ref{P-cylinder2a}) recovers the cell theorem [Eq.~(\ref{cell-theorem-cylinder})],
while evaluating at $r=0$ yields
\begin{equation}
P(R)=\frac{1}{R}\int_0^R du\, \left(\frac{|\psi'(u)|^2}{8\pi\lambda_B}
+n(u)-u n_f(u)\psi'(u)\right).
\label{P-cylinder2b}
\end{equation}
Interestingly, in cylindrical geometry, the bulk pressure can be expressed as an integral 
over the radial coordinate, but is not obtained by simply substituting $r$ for $x$ 
in the corresponding expression in planar geometry [Eq.~(\ref{P-plane2})].
Numerical evaluation confirms that Eqs.~(\ref{P-cylinder2a}) and (\ref{P-cylinder2b}) 
give exactly the same bulk pressure as Eq.~(\ref{cell-theorem-cylinder}) [Sec.~\ref{validation}].
For applications to swelling of ionic microgels [Sec.~\ref{ionic-microgels-cylindrical}], 
it is useful to relate the normal component of the electrostatic pressure profile 
[Eq.~(\ref{Pe-normal})] to the gel pressure:
\begin{equation}
P_{e,\parallel}(r)=n(R)-P_g(r)-\frac{1}{4\pi\lambda_B}\int_r^R du\, \frac{|\psi'(u)|^2}{u},
\label{Pe-cylinder2a}
\end{equation}
which follows from differentiating and then integrating Eq.~(\ref{P-cylinder2a})
with respect to $r$ and identifying the gel pressure
\begin{equation}
P_g(r)=-\int_r^R du\, n_f(u)\psi'(u),
\label{Pgr}
\end{equation}
the diagonal element $({\bf P}_g)_{rr}$ of the gel pressure tensor.

\begin{figure}
\includegraphics[width=0.9\columnwidth]{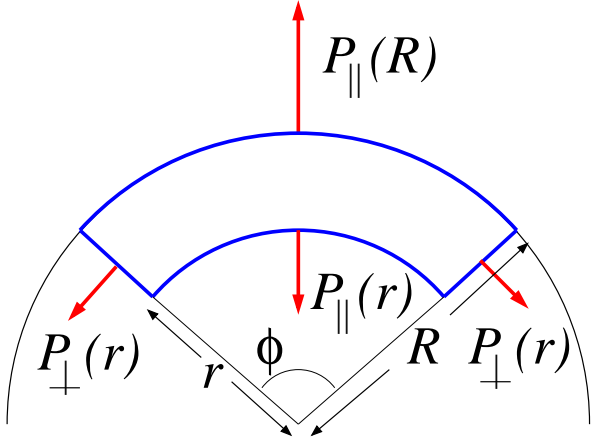} \\[2ex]
\vspace*{-0.2cm}
\caption{
Cylindrical sectors of a cylindrical cell, whose cross-sections are enclosed by circular arcs
of radius $R$ or $r$ with central angle $\phi$ and two straight radial segments. 
Mechanical equilibrium of the section highlighted in blue requires that the net force 
exerted by the normal pressures $P_{\parallel}$ and transverse pressures $P_{\perp}$ 
(red arrows) is zero.  This force balance condition is expressed by 
Eqs.~(\ref{P-cylinder2a})--(\ref{P-cylinder-force-balance}).
}\label{sector-fig}
\end{figure}

To further elucidate the significance of Eq.~(\ref{P-cylinder2a}), we now show that 
this expression follows also from the condition of mechanical equilibrium.
In fact, Eq.~(\ref{P-cylinder2a}) can be equivalently expressed in the form
\begin{equation}
RP(R)=rP_{\parallel}(r)+\int_r^R du\, P_{\bot}(u),
\label{P-cylinder-force-balance}
\end{equation}
where 
\begin{equation}
P_{\parallel}(r)=P_{e,\parallel}(r)+P_g(r)
=n(R)-\frac{1}{4\pi\lambda_B}\int_r^R du\, \frac{|\psi'(u)|^2}{u}
\label{P-normal-cylinder}
\end{equation}
is the normal component and
\begin{equation}
P_{\bot}(r)=P_{e,\bot}(r)+P_g(r)
\label{P-transverse}
\end{equation}
the transverse component of the total pressure tensor. Here
$P_{e,\parallel}(r)$ and $P_{e,\bot}(r)$ are the normal and transverse components
of the electrostatic pressure tensor [Eqs.~(\ref{Pe-normal}) and (\ref{Pe-transverse})]
and we assume uniform swelling and $P_g(R)=0$.
As illustrated in Fig.~\ref{sector-fig}, the left side of Eq.~(\ref{P-cylinder-force-balance})
is proportional to the magnitude of the net force acting on the cylindrical surface of a 
cylindrical sector (pie slice) of the cell whose cross-section is enclosed by a circular arc
of radius $R$ subtending a central angle $\phi$ and two straight radial segments. By symmetry, 
the net force points in the radial direction and is proportional to $R\sin(\phi/2)$. 
The first term on the right side of Eq.~(\ref{P-cylinder-force-balance}) is similarly 
proportional to the magnitude of the net force acting on the cylindrical surface of a 
cylindrical sector of radius $r$ subtending the same central angle, which is proportional to 
$r\sin(\phi/2)$. The second term on the right side is proportional to the magnitude 
of the net force acting on the flat sides of the sector. Integration along the radial coordinate 
is required, as the transverse pressure varies with $r$.
By symmetry, this net force also points in the radial direction and 
is proportional to $\sin(\phi/2)$. Thus, Eq.~(\ref{P-cylinder-force-balance}) simply 
expresses the condition that the net force on any sector of the cell vanishes.
Since the size and location of the sector are arbitrary, it follows that 
mechanical equilibrium prevails at every point in the cylinder.

The fact that precisely the same expression for the bulk pressure results from two
seemingly independent approaches -- a variational minimization of the free energy and 
a force balance condition involving the pressure tensor -- provides a strong 
consistency check on our calculations. 
In passing, we note that our expressions for $P_{\parallel}(r)$ and $P_{\bot}(r)$ 
(neither of which is spatially constant) satisfy the condition for mechanical equilibrium,
\begin{equation}
(\nabla\cdot{\bf P})_{\parallel}=\frac{\partial P_{\parallel}}{\partial r}
+\frac{1}{r}(P_{\parallel}-P_{\bot})=0,
\label{divergence-cylinder}
\end{equation}
which follows from substituting Eq.~(\ref{P-matrix}) into the general expression for 
the divergence of a second-order tensor in cylindrical polar coordinates~\cite{RHB,sochi2016}.
Furthermore, since $P_{\parallel}(r)$ 
and $P_{\bot}(r)$ are everywhere finite, Eq.~(\ref{divergence-cylinder})
implies that the equilibrium normal pressure is continuous.

\subsection{Spherical Cell Model}\label{spherical-cell-model}

In a spherically symmetric spherical cell of radius $R$, the Gibbs free energy functional 
now takes the form
\begin{equation}
G=-4\pi\int_0^R dr\, g(r,\psi,\psi'),
\label{G-sphere}
\end{equation}
where $r$ is the radial distance from the center and 
\begin{eqnarray}
g(r,\psi,\psi')&=&r^2\left(\frac{|\psi'(r)|^2}{8\pi\lambda_B}+2n_0\cosh\psi(r)\right.
\nonumber\\[1ex]
&+&\left.\frac{}{}n_f(r)\psi(r)-P(R)\right).
\label{g-sphere}
\end{eqnarray}
The geometric factor of $r^2$, from spherical symmetry in spherical geometry, further modifies 
the results derived in the planar and cylindrical cell models.
Substituting $g$ from Eq.~(\ref{g-sphere}) into Eqs.~(\ref{ELr})-(\ref{dg-dr1}) 
recovers the PB equation [Eq.~(\ref{PB})] in spherical polar coordinates,
the electroneutrality condition, $\psi'(R)=0$, and the cell theorem 
for the pressure [Eq.~(\ref{cell-theorem-cylinder})].
Integrating Eq.~(\ref{dg-dr2}) from $r$ to $r=R$ and using 
Eq.~(\ref{g-sphere}) to evaluate $\partial g/\partial r$, we have 
\begin{equation}
\left(g-\psi'\frac{\partial g}{\partial\psi'}\right){\Big |}_r^R=\int_r^R du\, \left(\frac{2g}{u}
+u^2 n_f'(u)\psi(u)\right).
\label{P-sphere1}
\end{equation}
Substituting Eq.~(\ref{g-sphere}) for $g$ on both sides of Eq.~(\ref{P-sphere1}) and 
integrating by parts, we obtain a general expression for the bulk pressure in the spherical cell model: 
\\[1ex]
\begin{widetext}
\begin{equation}
P(R)=\frac{r^2}{R^2}\left(-\frac{|\psi'(r)|^2}{8\pi\lambda_B}+n(r)\right)
+\frac{1}{R^2}\int_r^R du\, \left[2u\left(\frac{|\psi'(u)|^2}{8\pi\lambda_B}
+n(u)\right)-u^2 n_f(u)\psi'(u)\right].
\label{P-sphere2a}
\end{equation}
\end{widetext}
Evaluating Eq.~(\ref{P-sphere2a}) at $r=0$ yields
\begin{equation}
P(R)=\frac{1}{R^2}\int_0^R du\, \left[2u\left(\frac{|\psi'|^2}{8\pi\lambda_B}
+n\right)-u^2 n_f\psi'\right].
\label{P-sphere2b}
\end{equation}
Once again, the bulk pressure $P(R)$ can be expressed as an integral over the radial coordinate,
but is not obtained by simply substituting $r$ for $x$ in Eq.~(\ref{P-plane2}).
Applications to swelling of ionic microgels [Sec.~\ref{ionic-microgels-spherical}] are
facilitated by relating $P_{e,\parallel}(r)$ [Eq.~(\ref{Pe-normal})] to the gel pressure:
\begin{equation}
P_{e,\parallel}(r)=n(R)-P_g(r)-\frac{1}{2\pi\lambda_B}\int_r^R du\, \frac{|\psi'(u)|^2}{u},
\label{Pe-sphere2a}
\end{equation}
which follows from differentiating and then integrating Eq.~(\ref{P-sphere2a}).
The normal component of the total pressure in the spherical cell is then
\begin{equation}
P_{\parallel}(r)=P_{e,\parallel}(r)+P_g(r)
=n(R)-\frac{1}{2\pi\lambda_B}\int_r^R du\, \frac{|\psi'(u)|^2}{u}.
\label{Pr-sphere}
\end{equation}

As in the cylindrical cell model, we can validate our expression for the bulk pressure
[Eq.~(\ref{P-sphere2a})] by expressing it as a force balance condition on a spherical sector:
\begin{equation}
R^2P(R)=r^2P_{\parallel}(r)+2\int_r^R du\, uP_{\bot}(u),
\label{P-sphere-force-balance}
\end{equation}
where the left side is proportional to the magnitude of the net force acting on the spherical 
surface of a spherical sector of the cell whose cross-section is enclosed by a spherical cap
of radius $R$ and a cone with apex at the center of the sphere.
By symmetry, the net force is in the radial direction and is proportional to $R^2$. 
The first term on the right side is similarly proportional to the magnitude of the 
net force acting on the spherical surface of a spherical sector of radius $r$ subtending 
the same solid angle, which is proportional to $r^2$. The second term on the right side 
is proportional to the magnitude of the net force acting on the lateral surface of the cone, 
which by symmetry also points in the radial direction. A radial integration is required, 
as the transverse pressure varies with $r$. 
Finally, we note that our relations for $P_{\parallel}(r)$ and $P_{\bot}(r)$ satisfy 
the condition for mechanical equilibrium,
\begin{equation}
(\nabla\cdot{\bf P})_{\parallel}=\frac{\partial P_{\parallel}}{\partial r}
+\frac{2}{r}(P_{\parallel}-P_{\bot})=0,
\label{divergence-sphere}
\end{equation}
obtained by substituting Eq.~(\ref{P-matrix}) into the general expression for 
the divergence of a tensor in spherical polar coordinates~\cite{RHB,sochi2016}.
Note the factor of 2 [{\it cf}.~Eq.~(\ref{divergence-cylinder})].
As in cylindrical geometry, neither component of the pressure 
is spatially constant and $P_{\parallel}(r)$ is continuous.

The approach described above can be implemented for any fixed charge distribution.
Next we illustrate applications to idealized distributions that model microgels.

\section{Applications of PB Theory}\label{applications}

\subsection{Planar Ionic Microgels}\label{planar-gels}

The planar cell model has been widely used to model the structure of electrolytes near 
charged surfaces~\cite{Israelachvili1992,EvansWennerstrom}, e.g., metallic electrodes, 
charged colloids, and lipid membranes~\cite{Ben-Yaakov-Andelman-Harries-Podgornik-jpcb2009,
Ben-Yaakov-Andelman-Harries-Podgornik-jpcm2009,Markovich-Andelman-Podgornik2016}.
If the fixed charge uniformly coats a flat, impenetrable  wall at $x=a$ with valence/unit area
$\sigma$, then $n_f=\sigma\delta(x-a)$ and the electrostatic pressure is uniform throughout 
the available volume of the cell ($a< x < L$). According to Eqs.~(\ref{P-plane2}) and
(\ref{Pg-plane}), precisely at the surface ($x=a$), $P_{e,\parallel}(x)$ and $P_g(x)$ are 
discontinuous, jumping by equal but opposite values of magnitude $\sigma\psi'(a)$, 
such that the total pressure remains continuous. Evaluating Eq.~(\ref{Pe-plane1}) at $x=a$,
and invoking the boundary condition $\psi'(a)=4\pi\lambda_B\sigma$, yields the exact 
contact value theorem~\cite{Henderson-Blum1978,Henderson1979,wennerstrom1982,deserno-holm2001},
\begin{equation}
P(L)=n_+(a)+n_-(a)-2\pi\lambda_B\sigma^2,
\label{contact-value-theorem}
\end{equation}
which is valid only for planar charged surfaces.

The planar cell model can also be used, however, to model a flat, {\it permeable} membrane
or a slab of ionic gel. The latter we refer to as a planar microgel, since one dimension of
the gel is not macroscopic. If the fixed charge coats the surface of a permeable membrane or gel 
at $x=a$, then $P_e$ is discontinuous, jumping between different values on the left and right 
sides of the membrane. Evaluating Eq.~(\ref{Pe-plane1}) at $x=0$ and $x=L$, we have 
(with $n\equiv n_++n_-$)
\begin{equation}
P_{e,\parallel}(x)=
\left\{ \begin{array}
{l@{\quad}l}
2n_0\cosh\psi(0)=n(0),
& x\le a \\[2ex]
2n_0\cosh\psi(L)=n(L),
& x>a. \end{array} \right.
\label{delta-P-plane}
\end{equation}
Evaluating Eq.~(\ref{Pe-plane1}) as $x\to a$ from both sides,
\begin{equation}
P_{e,\parallel}(x)=
\left\{ \begin{array}
{l@{\quad}l}
n(a)-{\displaystyle\frac{1}{8\pi\lambda_B}
\lim\limits_{\delta\to 0}\psi'(a-\delta)^2},
& x\le a \\[2ex]
n(a)-{\displaystyle\frac{1}{8\pi\lambda_B}
\lim\limits_{\delta\to 0}\psi'(a+\delta)^2},
& x>a, \end{array} \right.
\label{Pe-flat-gel}
\end{equation}
which implies a discontinuity (osmotic pressure) of
\begin{equation}
\Delta P_e=\frac{1}{8\pi\lambda_B}
\lim\limits_{\delta\to 0}\left[\psi'(a+\delta)^2-\psi'(a-\delta)^2\right].
\label{delta-Pe-flat-gel1}
\end{equation}
Combining this result with the boundary condition
\begin{equation}
\lim_{\delta\to 0}\left[\psi'(a+\delta)-\psi'(a-\delta)\right]=4\pi\lambda_B\sigma
\label{membrane-bc}
\end{equation}
yields an alternative expression for the difference in electrostatic pressure across 
the membrane or gel surface:
\begin{equation}
\Delta P_e=\frac{\sigma}{2}\lim_{\delta\to 0}\left[\psi'(a+\delta)+\psi'(a-\delta)\right]
=n(0)-n(L).
\label{delta-Pe-flat-gel2}
\end{equation}
Because the electric field points away from the surface, the two terms on the right are
of opposite sign. In the symmetric case, $a=L/2$, the terms cancel and $\Delta P_e=0$.
Equations~(\ref{delta-Pe-flat-gel1}) and (\ref{delta-Pe-flat-gel2}) are equivalent forms
of a new ``contact value theorem" for surface-charged permeable membranes or planar microgels.

If the fixed charge is instead evenly distributed over the volume of a microgel in
the region $0\le x\le a$, then
\begin{equation}
n_f(x)=\frac{\sigma}{a}\theta(x-a),
\label{nf-volume-plane}
\end{equation}
where $\theta(x)$ is the unit step function. While fixed charges are discretely
distributed within a gel, thermal motion of polymer chains makes a continuum model reasonable. 
The electrostatic pressure is then given by Eq.~(\ref{Pe-plane2}) as
\begin{equation}
P_{e,\parallel}(x)=
\left\{ \begin{array}
{l@{\quad}l}
n(L)+\frac{\displaystyle\sigma}{\displaystyle a}\left[\psi(a)-\psi(x)\right],
& x\le a \\[2ex]
n(L),
& x>a, \end{array} \right.
\label{Pe-plane-gel}
\end{equation}
implying an internal electrostatic osmotic pressure
\begin{equation}
\Delta P_e(x)=\frac{\sigma}{a}\left[\psi(a)-\psi(x)\right].
\label{deltaPe-plane-gel}
\end{equation}
In contrast to the case of a surface-charged microgel, the electrostatic pressure of a
volume-charged microgel does not make a jump at the surface, but instead rises continuously 
with increasing distance into the gel.

\subsection{Cylindrical Ionic Microgels}\label{ionic-microgels-cylindrical}

The cylindrical cell model has been extensively used to model polyelectrolyte solutions and 
suspensions of charged rodlike colloidal particles~\cite{marcus1955,wennerstrom1982,deserno-holm2001}.
For reference, we first consider an impermeable cylinder of cross-sectional radius $a$ 
and fixed charge per unit length $-\lambda e$ uniformly spread over the surface 
with number density 
\begin{equation}
n_f(r)=\frac{\lambda}{2\pi a}\delta(r-a).
\label{nf-surface-cylinder}
\end{equation}
The bulk pressure is given by Eq.~(\ref{P-cylinder2a}) evaluated at $r=a$:
\begin{equation}
P(R)=\frac{a}{R}\left[n(a)-2\pi\lambda_B\sigma^2\right]
+\frac{1}{R}\int_a^R du\, \left(\frac{|\psi'|^2}{8\pi\lambda_B}+n\right),
\label{P-hard-cylinder}
\end{equation}
where $\sigma\equiv\lambda/(2\pi a)$ is the valence/unit area and we have applied 
the boundary condition for the electric field at the cylinder surface,
\begin{equation}
\psi'(a)=4\pi\lambda_B\sigma=2\lambda_B\lambda/a.
\label{cylinder-bc}
\end{equation}
Equation~(\ref{P-hard-cylinder}), derived here within PB theory, represents a 
contact value theorem for charged cylinders~\cite{wennerstrom1982}.
As noted in Sec.~\ref{cylindrical-cell-model}, this theorem can be interpreted 
as a force balance condition on a cylindrical sector of the cell.

Turning to a cylindrical microgel with uniform surface charge per unit length $-\lambda e$, 
the fixed charge density is the same as for an impermeable cylinder [Eq.~(\ref{nf-surface-cylinder})], 
but the surface is now {\it permeable} to microions. Substituting $n_f(r)$ into 
Eqs.~(\ref{Pe-cylinder2a}) and (\ref{Pgr}), we find that $P_{e,\parallel}(r)$ 
jumps discontinuously at $r=a$. 
The magnitude of this jump is the electrostatic osmotic pressure of the microgel:
\begin{equation}
\Delta P_e=-\Delta P_g
=\frac{\lambda}{4\pi a}\lim_{\delta\to 0}\left|\psi'(a+\delta)+\psi'(a-\delta)\right|.
\label{delta-P-mechanical-cylinder-surface-charge}
\end{equation}
Equation~(\ref{delta-P-mechanical-cylinder-surface-charge}) can be viewed as
a contact value theorem for cylindrical surface-charged ionic microgels.

It is important to note that, in contrast to planar geometry, the normal and transverse 
pressures are now not spatially uniform. Thus, whereas the electrostatic osmotic pressure 
of a planar surface-charged microgel is proportional to the difference in microion density 
between the inner ($x=0$) and outer ($x=L$) boundaries of the cell [Eq.~(\ref{delta-Pe-flat-gel2})],
the same relation does {\it not} hold for a cylindrical microgel.
The electrostatic osmotic pressure of a cylindrical surface-charged microgel is not 
simply proportional to the difference in microion density between the 
cylinder axis ($r=0$) and the edge of the cell ($r=R$), but rather equals the jump
in electrostatic pressure at the microgel surface ($r=a$).

Next, we consider a cylindrical microgel with fixed charge per unit length $-\lambda e$ 
uniformly distributed over its volume with number density 
\begin{equation}
n_f(r)=\frac{\lambda}{\pi a^2}\theta(r-a).
\label{nf-volume-cylinder}
\end{equation}
Substituting this form of $n_f(r)$ into Eq.~(\ref{Pgr}) yields
an internal electrostatic osmotic pressure (for $r<a$)
\begin{equation}
\Delta P_e(r)=P_{e,\parallel}(r)-P_{\parallel}(r)=-P_g(r)
=\frac{\lambda}{\pi a^2}\left[\psi(a)-\psi(r)\right],
\label{deltaPe-cylinder-gel}
\end{equation}
which is continuous at the surface of the gel and increases from the gel surface inward.

\subsection{Spherical Ionic Microgels}\label{ionic-microgels-spherical}

The spherical cell model is a well-established model of charge-stabilized colloidal 
suspensions~\cite{marcus1955,wennerstrom1982,deserno-holm2001}. For comparison,
we first consider an impermeable sphere, modeling a colloidal particle, 
of radius $a$ and uniform fixed surface charge $-Ze$ with number density 
\begin{equation}
n_f(r)=\frac{Z}{4\pi a^2}\delta(r-a).
\label{nf-surface-sphere}
\end{equation}
The bulk pressure is given by Eq.~(\ref{P-sphere2a}) evaluated at $r=a$:
\begin{equation}
P(R)=\frac{a^2}{R^2}\left[n(a)-2\pi\lambda_B\sigma^2\right]
+\frac{2}{R^2}\int_a^R du\, u\left(\frac{|\psi'|^2}{8\pi\lambda_B}+n\right),
\label{P-colloid}
\end{equation}
where $\sigma=Z/(4\pi a^2)$ is the valence/unit area and we have applied the 
boundary condition for the electric field at the particle surface,
\begin{equation}
\psi'(a)=4\pi\lambda_B\sigma=Z\lambda_B/a^2.
\label{sphere-bc}
\end{equation}
Equation~(\ref{P-colloid}) represents a contact value theorem 
for spherical charged colloids~\cite{wennerstrom1982}.
As discussed in Sec.~\ref{spherical-cell-model}, this theorem can be interpreted 
as a force balance condition on a spherical sector of the cell.

Turning to a spherical microgel with uniform surface charge $-Ze$, the fixed charge density 
is the same as for an impermeable spherical colloid [Eq.~(\ref{nf-surface-sphere})].
Substituting $n_f(r)$ into Eqs.~(\ref{Pe-sphere2a}) and (\ref{Pgr}), we find that 
$P_{e,\parallel}(r)$ jumps discontinuously at $r=a$. 
The magnitude of this jump is the electrostatic osmotic pressure of the microgel:
\begin{equation}
\Delta P_e=-\Delta P_g
=\frac{Z}{8\pi a^2}\lim_{\delta\to 0}\left|\psi'(a+\delta)+\psi'(a-\delta)\right|.
\label{delta-P-mechanical-microgel-surface-charge}
\end{equation}
Equation~(\ref{delta-P-mechanical-microgel-surface-charge}) can be regarded as 
a contact value theorem for spherical surface-charged ionic microgels.

As with cylindrical geometry, the normal and transverse pressures are not spatially uniform. 
Thus, the electrostatic osmotic pressure of a spherical surface-charged microgel is 
{\it not} simply proportional to the difference in microion density between the center ($r=0$)
and edge ($r=R$) of the cell, but rather equals the jump in electrostatic pressure 
at the microgel surface ($r=a$).
Our approach thus differs sharply from that of ref.~\cite{gasser2019},
in which the pressure inside the microgel is approximated as uniform and the
electrostatic osmotic pressure is claimed to depend only on the microion density at the center.

A spherical microgel with fixed charge $-Ze$ uniformly distributed over its volume
has charge number density 
\begin{equation}
n_f(r)=\frac{3Z}{4\pi a^3}\theta(r-a).
\label{nf-volume-sphere}
\end{equation}
Substituting this form of $n_f(r)$ into Eq.~(\ref{Pgr}) yields
an internal electrostatic osmotic pressure (for $r<a$)
\begin{equation}
\Delta P_e(r)=P_{e,\parallel}(r)-P_{\parallel}(r)=-P_g(r)
=\frac{3Z}{4\pi a^3}\left[\psi(a)-\psi(r)\right],
\label{deltaPe-sphere-gel}
\end{equation}
which, as in planar and cylindrical geometries, rises continuously from the surface inward.
To check and compare with the above results, obtained within PB theory, we next derive theorems 
that provide exact relations for the electrostatic component of the osmotic pressure 
in the cell model, which are not tied to PB theory.

\section{Exact Relations in the Cell Model}\label{exact}

\subsection{Osmotic Pressure of Ionic Microgels}\label{osmotic-pressures}
For comparison with the predictions of Poisson-Boltzmann theory, we next derive some exact 
statistical mechanical relations for the electrostatic contribution to the osmotic pressure 
of an ionic microgel in the cell model, following the approach developed in previous 
work~\cite{denton-tang2016} and inspired by Wennerstr\"om \etalia\cite{wennerstrom1982}.
The derivations start from a decomposition of the electrostatic part of the Hamiltonian, 
\begin{equation}
H_e=U_m(a)+U_{m\mu}(\{{\bf r}\};a)+U_{\mu\mu}(\{{\bf r}\}),
\label{He}
\end{equation}
into the microgel self-energy $U_m(a)$ and microgel-microion and microion-microion 
interaction energies, $U_{m\mu}(\{{\bf r}\};a)$ and $U_{\mu\mu}(\{{\bf r}\})$, respectively,
which depend on the coordinates of all $N$ microions, 
$\{{\bf r}\}\equiv\{{\bf r}_1,\ldots,{\bf r}_N\}$.
Only the first two terms in Eq.~(\ref{He}) depend on the microgel size.  
The microgel-microion interaction energy depends on the microgel-microion 
pair potential $v_{m\mu}({\bf r};a)$:
\begin{equation}
U_{m\mu}(\{{\bf r}\};a)=\sum_{i=1}^N v_{m\mu}({\bf r}_i;a).
\label{Ummu}
\end{equation}
For a suspension that is free to exchange microions with a salt reservoir, 
the electrostatic part of the semi-grand potential can be expressed as $\Omega=-\ln\Xi$, 
where (continuing to express energies in $k_BT$ units)
\begin{equation}
\Xi(\mu_0,V,T)~\propto~\sum_{N=0}^{\infty}\frac{e^{\mu_0 N}}{N!}
\prod_{i=1}^N \int_V d{\bf r}_i\, e^{-H_e}
\label{Xi}
\end{equation}
is the electrostatic partition function and $\mu_0$ is the chemical potential of microions 
in the reservoir.  The osmotic pressure of the suspension -- the difference in pressure 
between the suspension and reservoir -- is defined via the derivative of the free energy 
with respect to volume $V$. In a planar cell of constant surface area $A$ ($V=AL$), 
the bulk osmotic pressure is given by
\begin{equation}
P=\frac{1}{A}\frac{\partial}{\partial L}\ln\Xi.
\label{p0-planar}
\end{equation}
In a cylindrical cell of radius $R$ and height $h$ ($V=\pi R^2 h$),
\begin{equation}
P=\frac{1}{2\pi R h}\frac{\partial}{\partial R}\ln\Xi,
\label{p0-cylindrical}
\end{equation}
while in a spherical cell of radius $R$ ($V=4\pi R^3/3$),
\begin{equation}
P=\frac{1}{4\pi R^2}\frac{\partial}{\partial R}\ln\Xi.
\label{p0-spherical}
\end{equation}
Substituting for $\Xi$ from Eqs.~(\ref{He})-(\ref{Xi}) yields the respective cell theorems,
namely, Eqs.~(\ref{cell-theorem-plane}) and 
(\ref{cell-theorem-cylinder})~\cite{marcus1955,wennerstrom1982}. 

Similarly, the osmotic pressure of a permeable microgel is defined as the change in 
pressure at the surface of the gel. The electrostatic component of the osmotic pressure 
of an ionic microgel, $\Delta P_e$, can be defined via a derivative of the electrostatic 
semi-grand potential with respect to the gel volume $v$. In a planar cell with $v=Aa$,
\begin{equation}
\Delta P_e=\frac{\partial}{\partial v}\ln\Xi
=-\frac{1}{A}\left(\frac{\partial}{\partial a}U_m
+\la\frac{\partial}{\partial a}U_{m\mu}\ra\right),
\label{p1-planar}
\end{equation}
where angular brackets denote an ensemble average over microion configurations. 
In a cylindrical cell ($v=\pi a^2 h$),
\begin{equation}
\Delta P_e=-\frac{1}{2\pi a h}\left(\frac{\partial}{\partial a}U_m
+\la\frac{\partial}{\partial a}U_{m\mu}\ra\right),
\label{p1-cylindrical}
\end{equation}
while in a spherical cell ($v=4\pi a^3/3$),
\begin{equation}
\Delta P_e=-\frac{1}{4\pi a^2}\left(\frac{\partial}{\partial a}U_m
+\la\frac{\partial}{\partial a}U_{m\mu}\ra\right).
\label{p1-spherical}
\end{equation}
We emphasize that Eqs.~(\ref{p1-planar})-(\ref{p1-spherical}) yield 
average osmotic pressures, in contrast to the spatially varying pressure profiles 
derived from PB theory. Within the planar, cylindrical, and spherical cell models, 
these theorems for the electrostatic part of the osmotic pressure of
a permeable ionic microgel are formally exact. 
Next, we present explicit results for specific fixed charge distributions.

\subsection{Planar Ionic Microgels}\label{planar-cell-model-exact}

A slab of permeable ionic gel of thickness $a$, or a flat membrane at position $x=a$, 
with a surface charge/unit area $-\sigma e$, modeled by fixed charge number density 
$n_f(x)=\sigma\delta(x-a)$, has a self energy that is independent of $a$.
Since the electric field magnitude of the fixed charge, $|E_f|=2\pi\sigma e/\epsilon$, 
is constant, the fixed charge interacts with the microions with energy (in $k_BT$ units)
\begin{equation}
U_{m\mu}=2\pi\lambda_B\sigma\left(\sum_{i\, (x_i\le a)}z_i(a-x_i)+\sum_{i\, (x_i>a)}z_i(x_i-a)\right),
\label{Ummu-planar-surface}
\end{equation}
where $z_i=\pm 1$ is the valence of microion $i$ and the sums are restricted to microions
that are inside ($x_i\le a$) or outside ($x_i>a$) the gel.
Substituting into Eq.~(\ref{p1-planar}),
\begin{equation}
\Delta P_e=4\pi\lambda_B\sigma\left(\frac{\sigma}{2}-\la N_+\ra+\la N_-\ra\right),
\label{cell-theorem-planar-gel-surface-charge}
\end{equation}
where, for microion number density profiles $n_{\pm}(x)$, 
\begin{equation}
\la N_{\pm}\ra=\int_0^a dx\, n_{\pm}(x)
\label{Npm-planar}
\end{equation}
are mean numbers of counterions/coions per unit area inside the gel
and we used the electroneutrality condition
\begin{equation}
\sigma A=\sum_{i\, (x_i\le a)}z_i+\sum_{i\, (x_i>a)}z_i.
\label{electroneutrality-planar}
\end{equation}
Note that in the symmetric case of a membrane fixed at the center of the cell ($a=L/2$),
$\la N_+\ra-\la N_-\ra=\sigma/2$ and then $\Delta P_e=0$.

A slab of ionic gel with the same charge evenly distributed over its volume has a 
fixed charge number density $n_f(x)=(\sigma/a)\theta(x-a)$, generating an electric field
\begin{equation}
E_f(x)=
\left\{ \begin{array}
{l@{\quad}l}
{\displaystyle -\frac{4\pi\sigma e}{\epsilon}\left(\frac{x}{a}-\frac{1}{2}\right)},
& x\le a \\[2ex]
{\displaystyle -\frac{2\pi\sigma e}{\epsilon}},
& x>a. \end{array} \right.
\label{E-planar-volume}
\end{equation}
which stores a self energy (in $k_BT$ units)
\begin{equation}
U_m=\frac{A\epsilon}{8\pi k_BT}\int_0^L dx\, |E_f(x)|^2
=A\frac{\pi\lambda_B\sigma^2}{3}\left(\frac{3L}{2}-a\right).
\label{Um-planar-volume}
\end{equation}
The corresponding interaction energy between the fixed charge and the microions is
\begin{equation}
U_{m\mu}=2\pi\lambda_B\sigma\left(\sum_{i\, (x_i\le a)}z_i\left(\frac{x_i^2}{a}-x_i\right)
+\sum_{i\, (x_i>a)}z_i(x_i-a)\right)
\label{Ummu-planar-volume}
\end{equation}
plus a constant $U_0(a)$, dependent on $a$, determined by the choice of reference point.
Substituting Eqs.~(\ref{Um-planar-volume}) and (\ref{Ummu-planar-volume}) into 
Eq.~(\ref{p1-planar}) and choosing $U_0(a)=\pi\lambda_B\sigma^2 a$ yields
\begin{equation}
\Delta P_e=2\pi\lambda_B\sigma\left(\frac{2}{3}\sigma
-\la N_+\ra+\la N_-\ra+\frac{\la x^2\ra_+-\la x^2\ra_-}{a^2}\right),
\label{cell-theorem-planar-gel-volume-charge}
\end{equation}
where
\begin{equation}
\la x^2\ra_{\pm}=\int_0^a dx\, x^2n_{\pm}(x)
\label{x2pm-planar}
\end{equation}
are second moments of the microion density profiles inside the gel. Our choice of 
$U_0(a)$ ensures that Eqs.~(\ref{cell-theorem-planar-gel-surface-charge}) and
(\ref{cell-theorem-planar-gel-volume-charge}) agree in the thin-gel ($a\to 0$) and
high-charge ($\sigma\to\infty$) limits, in which case $n_+(x)\to(\sigma/a)\theta(x-a)$.
Of the six systems considered -- surface- and volume-charged planar, cylindrical, and spherical 
microgels -- the volume-charged planar microgel is the only one for which the reference potential 
affects the electrostatic pressure. In passing, we note that in the limit of a macroscopic gel,
much thicker than the Debye screening length ($\kappa a\gg 1$), when virtually all counterions 
are confined to the gel, the gel is electroneutral, $n_+(x)=n_f(x)$, and $\Delta P_e=0$.

\subsection{Cylindrical Ionic Microgels}\label{cylindrical-cell-model-exact}

For a cylindrical microgel with fixed charge uniformly distributed over its surface 
[Eq.~(\ref{nf-surface-cylinder})], the self-energy and the microgel-microion 
interaction energy per unit length are (in $k_BT$ units and to within $a$-independent constants)
\begin{equation}
U_m=-\lambda^2\lambda_B\ln\left(\frac{a}{R}\right)
\label{Um-cylindrical-surface}
\end{equation}
and 
\begin{equation}
U_{m\mu}=2\lambda\lambda_B\ln\left(\frac{a}{R}\right)\sum_{i\, (r_i\le a)} z_i. 
\label{vmmugel-cylindrical-surface}
\end{equation}
Substituting Eqs.~(\ref{Um-cylindrical-surface}) and (\ref{vmmugel-cylindrical-surface})   
into Eq.~(\ref{p1-cylindrical}) yields
\begin{equation}
\Delta P_e=\frac{\lambda\lambda_B}{\pi a^2}\left(\frac{\lambda}{2}-\la N_+\ra+\la N_-\ra\right),
\label{cell-theorem-microgel-cylinder-surface-charge}
\end{equation}
where, for microion radial number density profiles $n_{\pm}(r)$, 
\begin{equation}
\la N_{\pm}\ra=2\pi\int_0^a dr\, r n_{\pm}(r)
\label{Npm-cylindrical}
\end{equation}
are mean numbers of counterions/coions per unit length inside the microgel.

A cylindrical microgel with fixed charge uniformly spread over its volume
[Eq.~(\ref{nf-volume-cylinder})] creates an electric field
\begin{equation}
E_f(r)=
\left\{ \begin{array}
{l@{\quad}l}
{\displaystyle -\frac{2\lambda e}{\epsilon a^2}}r,
& r\le a \\[2ex]
{\displaystyle -\frac{2\lambda e}{\epsilon r}},
& r>a, \end{array} \right.
\label{E-cylindrical-volume}
\end{equation}
storing a self-energy per unit length (in $k_BT$ units)
\begin{equation}
U_m=\frac{\epsilon}{4k_BT}\int_0^R dr\, r|E_f(r)|^2
=\lambda^2\lambda_B\left[\frac{1}{4}-\ln\left(\frac{a}{R}\right)\right],
\label{Um-cylindrical-volume}
\end{equation}
and interacts with the microions with energy 
\begin{equation}
U_{m\mu}=2\lambda\lambda_B\sum_{i~(r_i\le a)} z_i\left[\ln\left(\frac{a}{R}\right)
+\frac{r_i^2}{2a^2}\right].
\label{vmmugel-cylindrical-volume}
\end{equation}
Substituting Eqs.~(\ref{Um-cylindrical-volume}) and (\ref{vmmugel-cylindrical-volume})   
into Eq.~(\ref{p1-cylindrical}) yields
\begin{equation}
\Delta P_e=\frac{\lambda\lambda_B}{\pi a^2}\left(\frac{\lambda}{2}-\la N_+\ra+\la N_-\ra 
+\frac{\la r^2\ra_+ - \la r^2\ra_-}{a^2}\right),
\label{cell-theorem-microgel-cylinder-volume-charge}
\end{equation}
where
\begin{equation}
\la r^2\ra_{\pm}=2\pi \int_0^a dr\, r^3 n_{\pm}(r)
\label{r2pm-cylindrical}
\end{equation}
are second moments of $n_{\pm}(r)$ inside the microgel.

\subsection{Spherical Ionic Microgels}\label{spherical-cell-model-exact}

For a spherical microgel whose fixed charge is uniformly spread over its surface, with 
charge number density described by Eq.~(\ref{nf-surface-sphere}), the self-energy and 
microgel-microion interaction energy are 
\begin{equation}
U_m=\frac{Z^2\lambda_B}{2a}
\label{Um-spherical-surface}
\end{equation}
and (to within a constant, independent of $a$)
\begin{equation}
U_{m\mu}= -\frac{Z\lambda_B}{a}\sum_{i\, (r_i\le a)}z_i.
\label{vmmugel-spherical-surface}
\end{equation}
Substituting Eqs.~(\ref{Um-spherical-surface}) and (\ref{vmmugel-spherical-surface})   
into Eq.~(\ref{p1-spherical}) yields
\begin{equation}
\Delta P_e=\frac{Z\lambda_B}{4\pi a^4}\left(\frac{Z}{2}-\la N_+\ra+\la N_-\ra\right),
\label{cell-theorem-microgel-surface-charge}
\end{equation}
where, for microion radial number density profiles $n_{\pm}(r)$, 
\begin{equation}
\la N_{\pm}\ra=4\pi\int_0^a dr\, r^2 n_{\pm}(r)
\label{Npm-spherical}
\end{equation}
are mean counterion/coion numbers inside the microgel.

For a spherical microgel with fixed charge uniformly distributed over its volume
[Eq.~(\ref{nf-volume-sphere})], the self-energy is 
\begin{equation}
U_m=\frac{3Z^2\lambda_B}{5a}
\label{Um-spherical-volume}
\end{equation}
and the microgel-microion interaction energy is 
\begin{equation}
U_{m\mu}=-\frac{Z\lambda_B}{2a}\sum_{i\, (r_i\le a)}
z_i\left(3-\frac{r_i^2}{a^2}\right).
\label{vmmugel-spherical-volume}
\end{equation}
Substituting 
into Eq.~(\ref{p1-spherical}) yields (see ref.~\cite{denton-tang2016})
\begin{equation}
\Delta P_e=\frac{3Z\lambda_B}{8\pi a^4}
\left( \frac{2}{5}Z - \la N_+\ra + \la N_-\ra
+ \frac{\la r^2\ra_+-\la r^2\ra_-}{a^2} \right),
\label{cell-theorem-microgel-volume-charge}
\end{equation}
where
\begin{equation}
\la r^2\ra_{\pm}=4\pi \int_0^a dr\, r^4 n_{\pm}(r)
\label{r2pm-spherical}
\end{equation}
are second moments of $n_{\pm}(r)$ inside the microgel.

The expressions derived above for the electrostatic component of the osmotic pressure 
of ionic microgels are exact within the cell model, independent of PB theory,
as is the cell theorem for the pressure outside~\cite{wennerstrom1982}.
These expressions may be combined with any theory of the counteracting gel component $\Delta P_g$ 
of the pressure to predict equilibrium swelling behavior by determining the radial
swelling ratio for which the total osmotic pressure vanishes. Recently, we analyzed 
swelling of ionic microgels~\cite{denton-tang2016,weyer-denton2018} 
by applying the Flory-Rehner theory of polymer 
networks~\cite{flory1953,flory-rehner1943-I,flory-rehner1943-II}.
The Flory-Rehner theory predicts
\begin{eqnarray}
\Delta P_g\frac{4\pi}{3}a^3 = &-&N_m\left[\alpha^3\ln\left(1-\frac{1}{\alpha^3}\right) +
\frac{\chi}{\alpha^3}+1\right] 
\nonumber\\[1ex]
&-& N_{\rm ch}\left(\alpha^2 - \frac{1}{2}\right),
\label{Flory-p}
\end{eqnarray}
where $\alpha\equiv a/a_0$ is the radial swelling ratio, defined as the ratio of the 
swollen radius to the collapsed radius, $N_m$ and $N_{\rm ch}$ are the numbers of monomers 
and chains per microgel, and $\chi$ is the Flory solvency parameter.
A similar approach can be applied to ionic microgels with any given distribution of fixed charge.
Alternatively, our general expressions for the electrostatic pressure may be coupled with 
more accurate microscopic models of polymer networks. 
Next we demonstrate that the exact statistical mechanical relations for the electrostatic pressure 
confirm the relations derived from PB theory in Sec.~\ref{applications}.

\section{Validation and Discussion}\label{validation} 

\subsection{Planar Ionic Microgels}\label{validation-planar-gels}

To validate the various expressions derived in Secs.~\ref{applications} and \ref{exact} 
for the electrostatic component of the pressure of a permeable ionic microgel, 
we solved the PB equation in the cell model in planar, cylindrical, and spherical geometries. 
For ionic microgels with planar symmetry, we solved Eq.~(\ref{PB}) in Cartesian coordinates, 
\begin{equation}
\psi''(x)=\kappa^2\sinh\psi(x)+4\pi\lambda_B n_f(x),
\label{PB-planar}
\end{equation}
with boundary conditions $\psi'(0)=\psi'(L)=0$ (imposing electroneutrality). 
For a surface-charged microgel or membrane, computing $P_e(x)$ from Eqs.~(\ref{delta-P-plane}) 
and (\ref{delta-Pe-flat-gel2}) confirms precise agreement with the corresponding exact result 
for the electrostatic osmotic pressure [Eq.~(\ref{cell-theorem-planar-gel-surface-charge})] 
when the various expressions are evaluated using the microion density profiles $n_{\pm}(x)$ 
resulting from solving Eq.~(\ref{PB-planar}).

For a volume-charged ionic microgel, evaluating Eq.~(\ref{deltaPe-plane-gel}) yields a continuously 
varying electrostatic pressure profile $P_e(x)$. Interestingly, the value at the left wall, $P_e(0)$, 
exceeds the exact electrostatic osmotic pressure [Eq.~(\ref{cell-theorem-planar-gel-volume-charge})].
Remarkably, however, the {\it average} of $\Delta P_e(x)$ 
[Eq.~(\ref{deltaPe-plane-gel})] over the gel width (or volume),
\begin{equation}
\la \Delta P_e\ra=
\frac{\sigma}{a^2}\int_0^a dx\, \left[\psi(a)-\psi(x)\right],
\label{Pe-plane-gel-average}
\end{equation}
precisely agrees with Eq.~(\ref{cell-theorem-planar-gel-volume-charge}), with the
appropriate choice of the microgel-microion reference potential $U_0(a)$. We conclude, 
therefore, that the exact result for the electrostatic component of the osmotic pressure 
inside a slab of ionic gel with uniformly distributed volume charge equals the average 
of the electrostatic osmotic pressure profile predicted by PB theory in the planar cell model.

\begin{figure}[t!]
\includegraphics[width=\columnwidth]{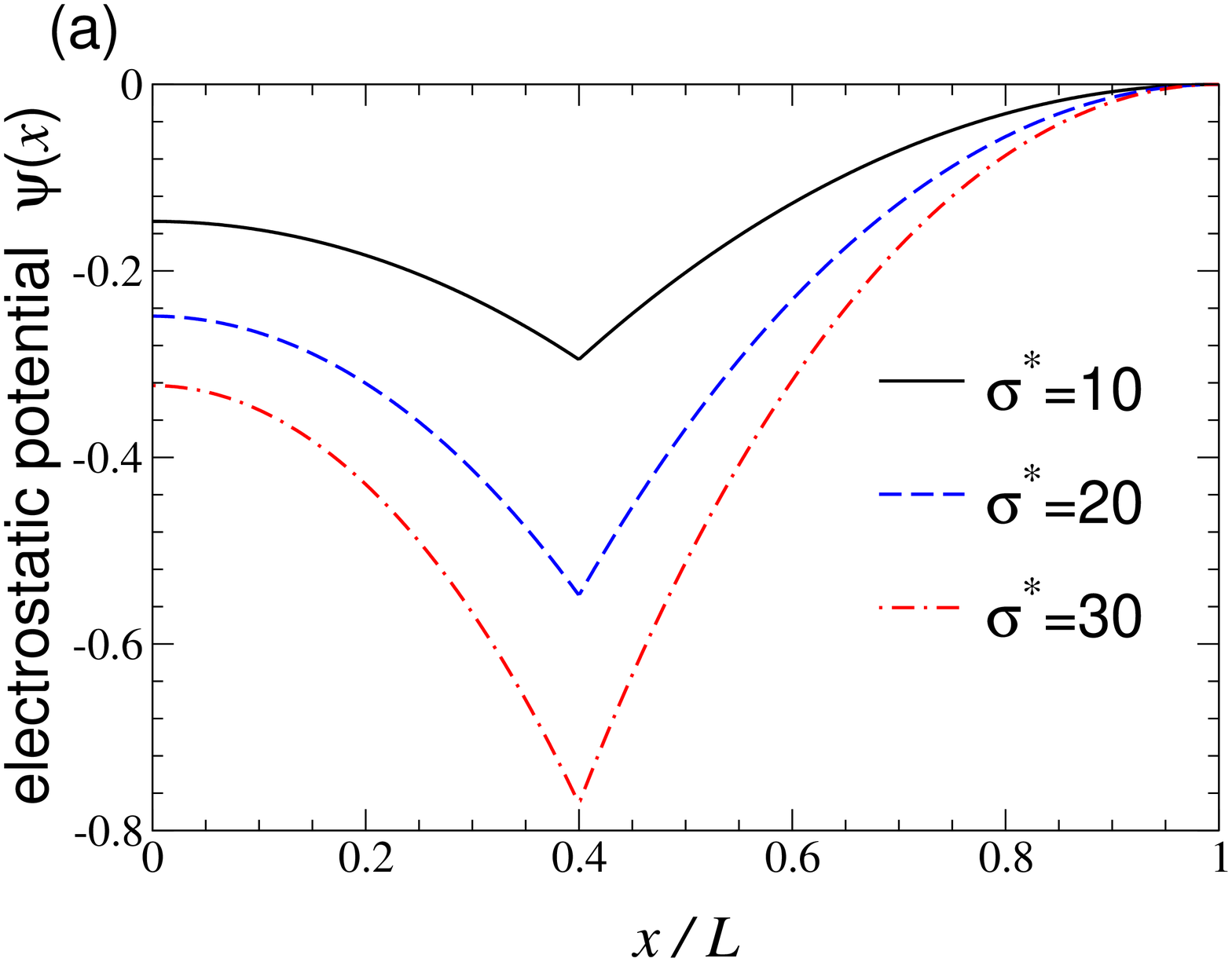}
\includegraphics[width=\columnwidth]{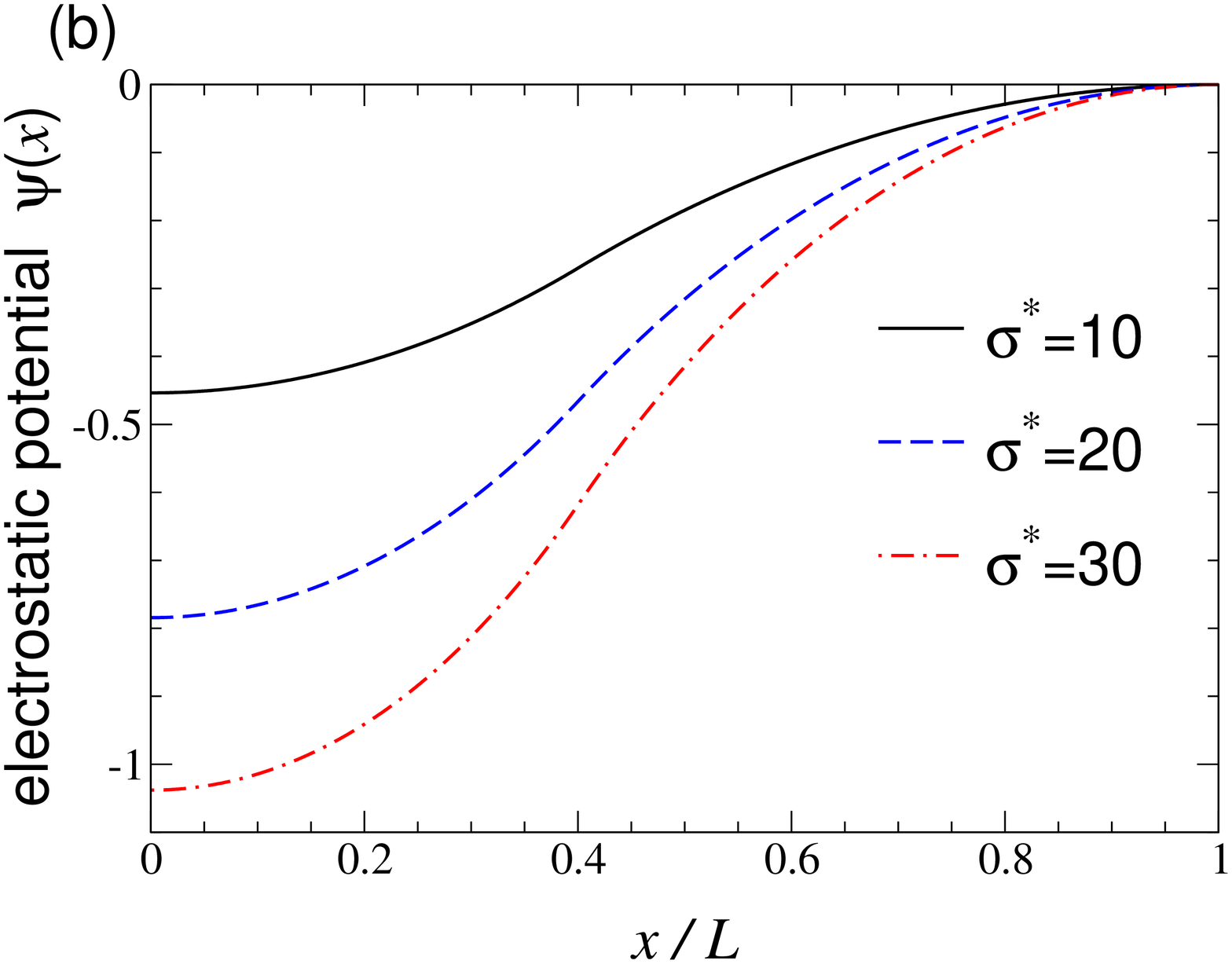}
\vspace*{-0.2cm}
\caption{
Reduced electrostatic potential $\psi(x)$ (shifted to 0 at $x=L$) vs.~distance $x$ from left wall 
in a planar cell of width $L=50$ nm for (a) surface-charged, (b) volume-charged planar microgels 
of swollen thickness $a=20$ nm and reduced valence/unit area $\sigma^*\equiv\sigma L^2=10$, 
20, 30 (solid, dashed, dash-dotted curves) at $T=293$ K in aqueous solution 
($\lambda_B=0.714$ nm) with negligible salt.
}\label{psix}
\end{figure}
%
\begin{figure}[t!]
\includegraphics[width=\columnwidth]{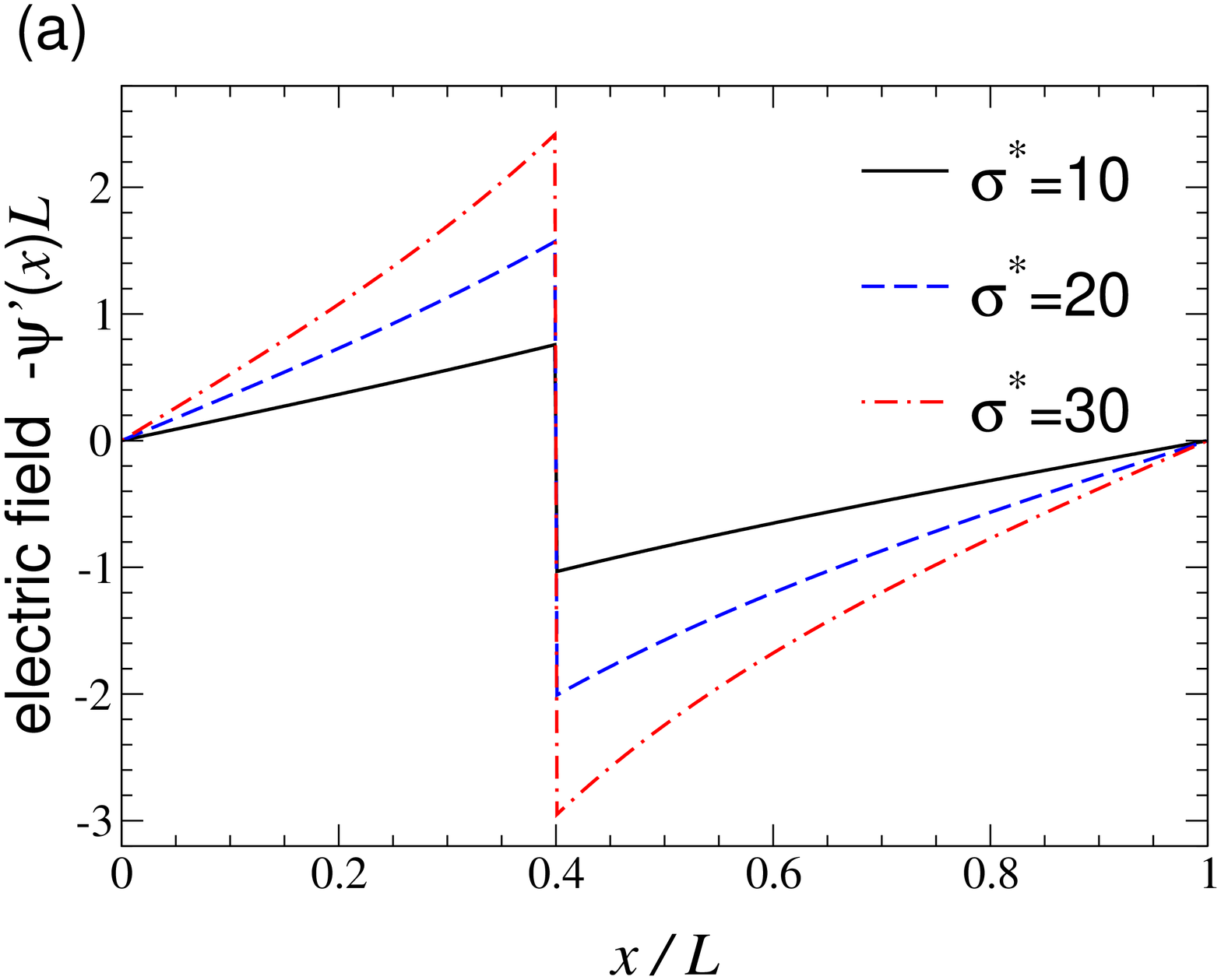}
\includegraphics[width=\columnwidth]{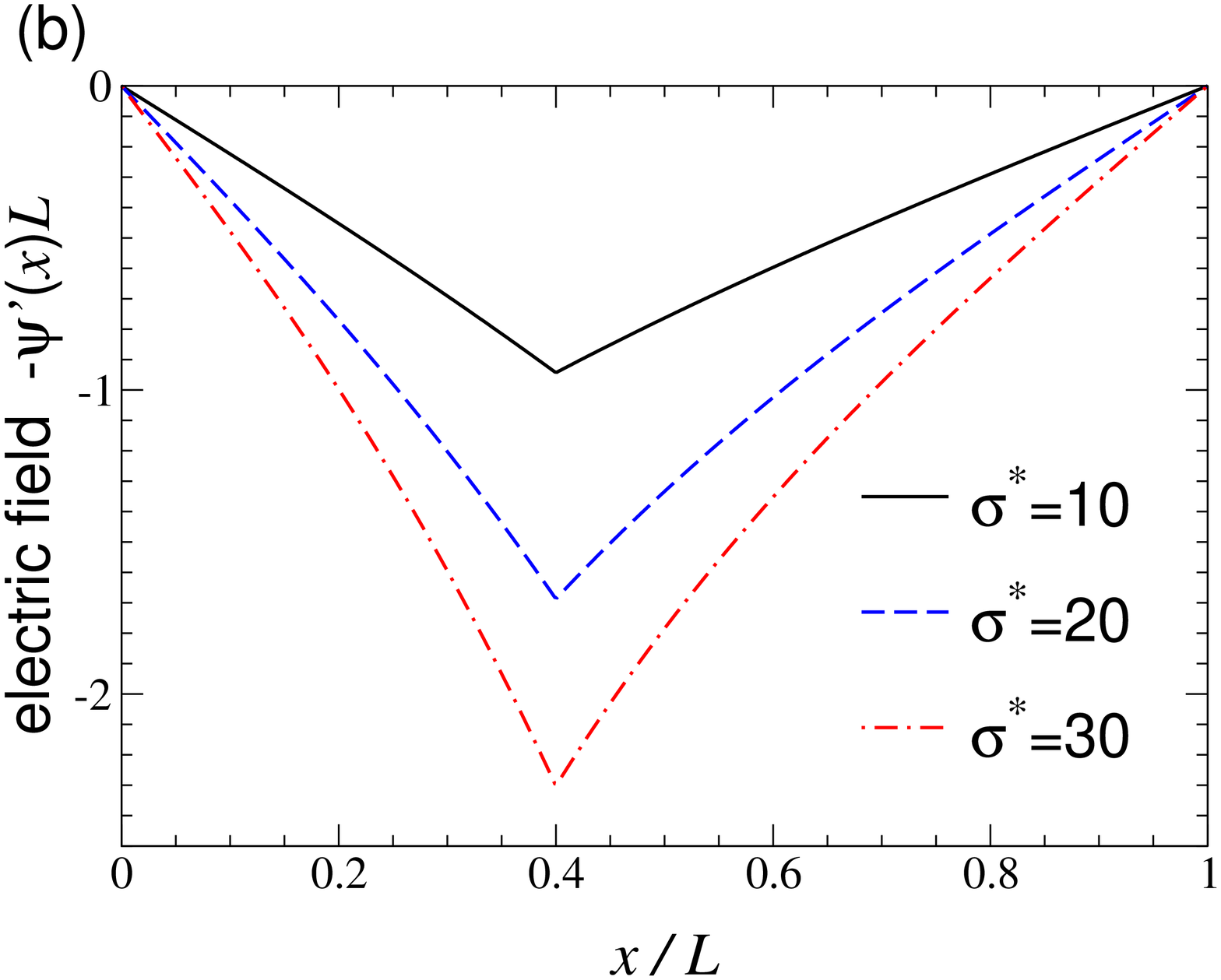}
\vspace*{-0.2cm}
\caption{
Reduced electric field $-\psi'(x)L$ vs.~distance $x$ from left wall of a planar cell for 
(a) surface-charged, (b) volume-charged planar microgels with system parameters of Fig.~\ref{psix}.
}\label{Ex}
\end{figure}

\begin{figure}[t]
\includegraphics[width=\columnwidth]{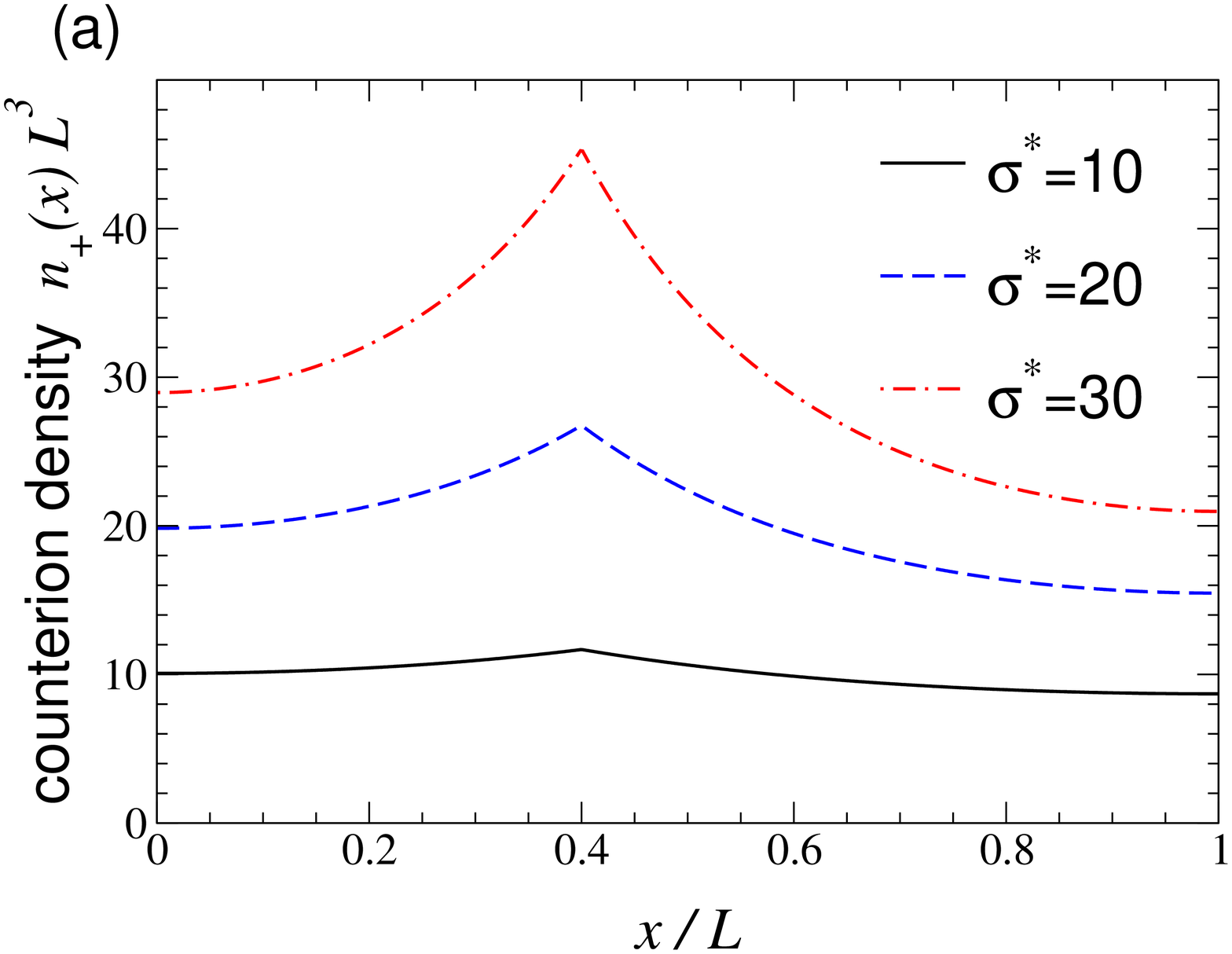}
\includegraphics[width=\columnwidth]{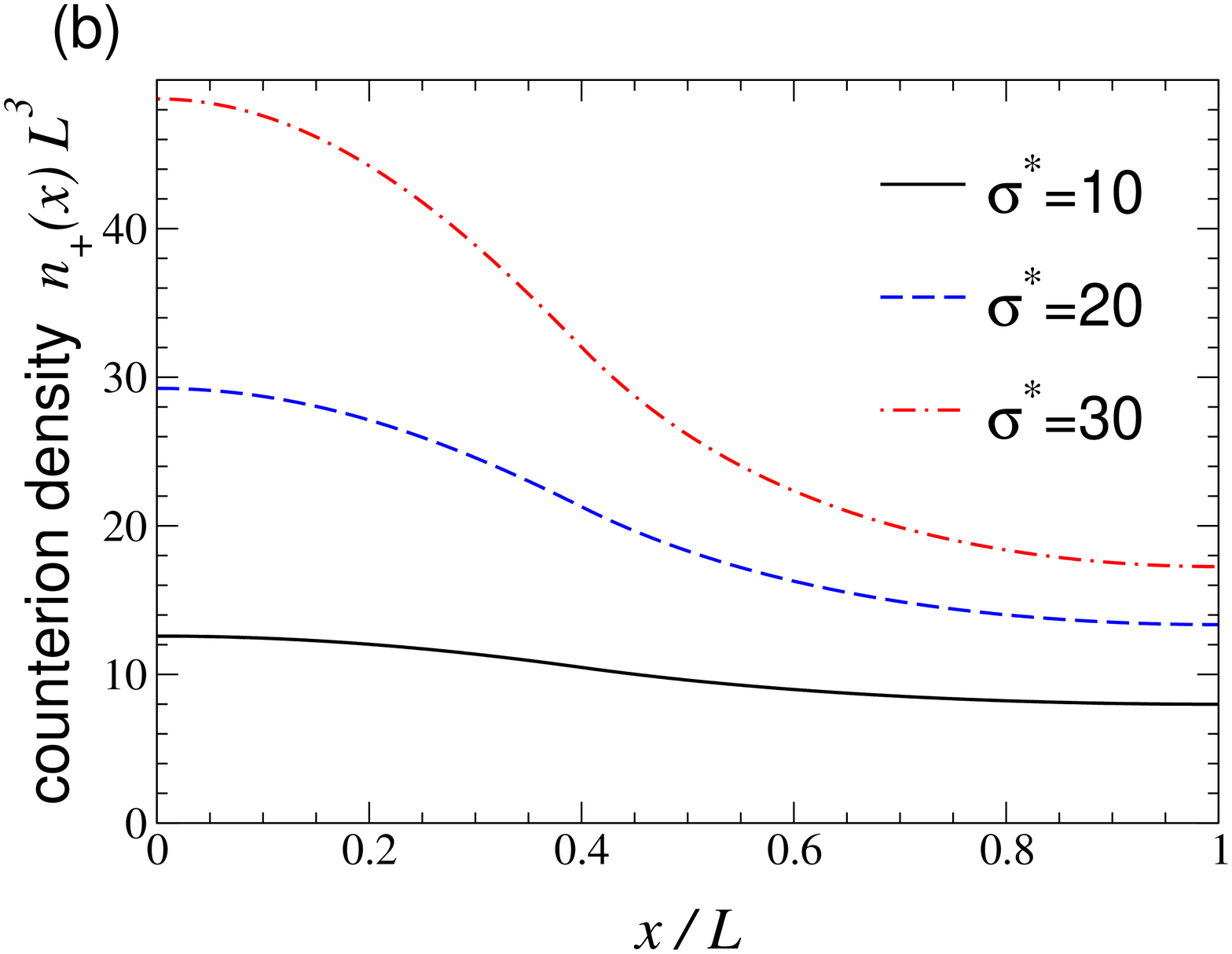}
\vspace*{-0.2cm}
\caption{
Counterion density $n_+(x)$ vs.~distance $x$ from left wall in the planar cell model for 
(a) surface-charged, (b) volume-charged planar microgels with system parameters of Fig.~\ref{psix}.
}\label{ncx}
\end{figure}

\begin{figure}[t]
\includegraphics[width=\columnwidth]{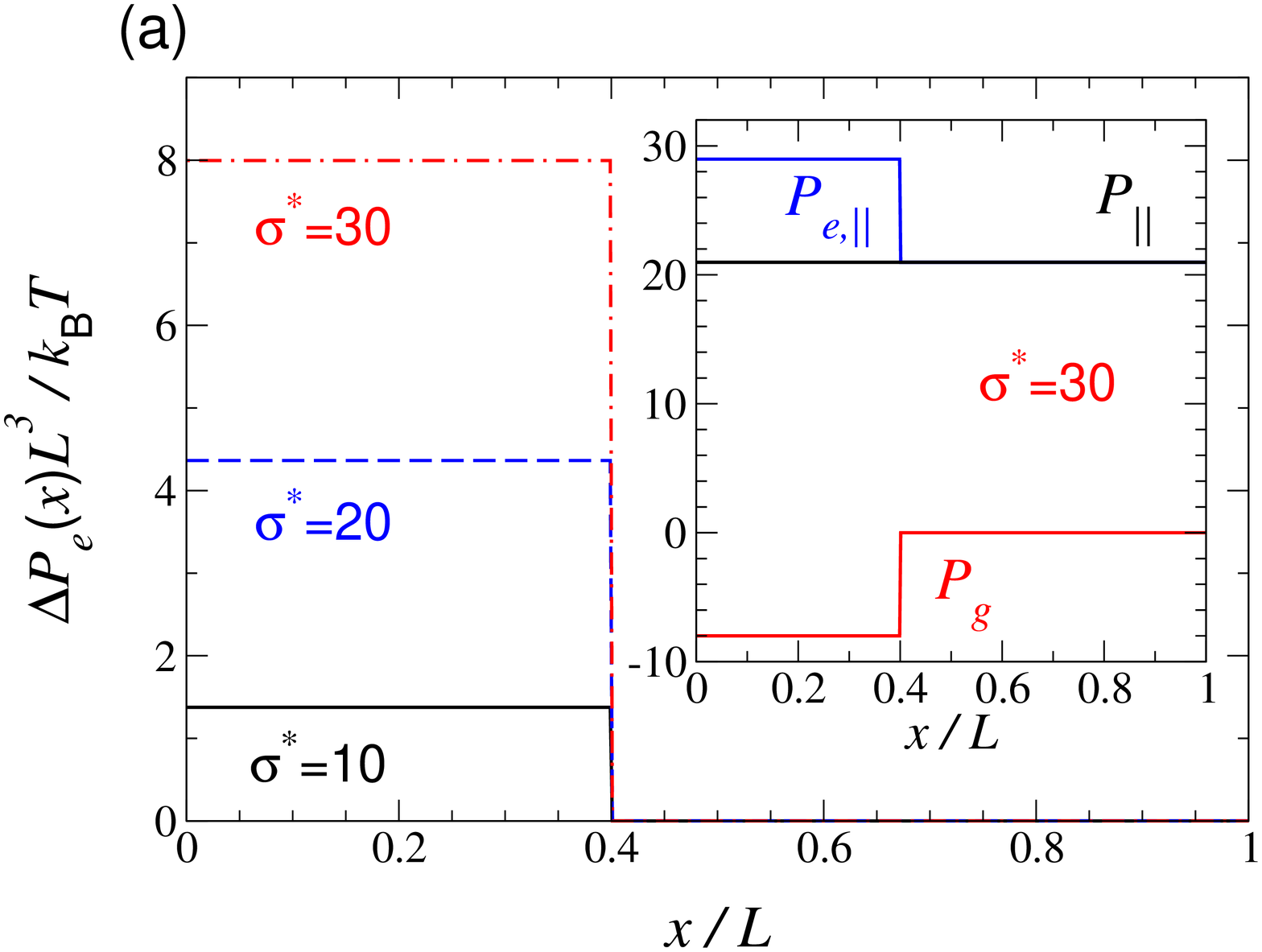}
\includegraphics[width=\columnwidth]{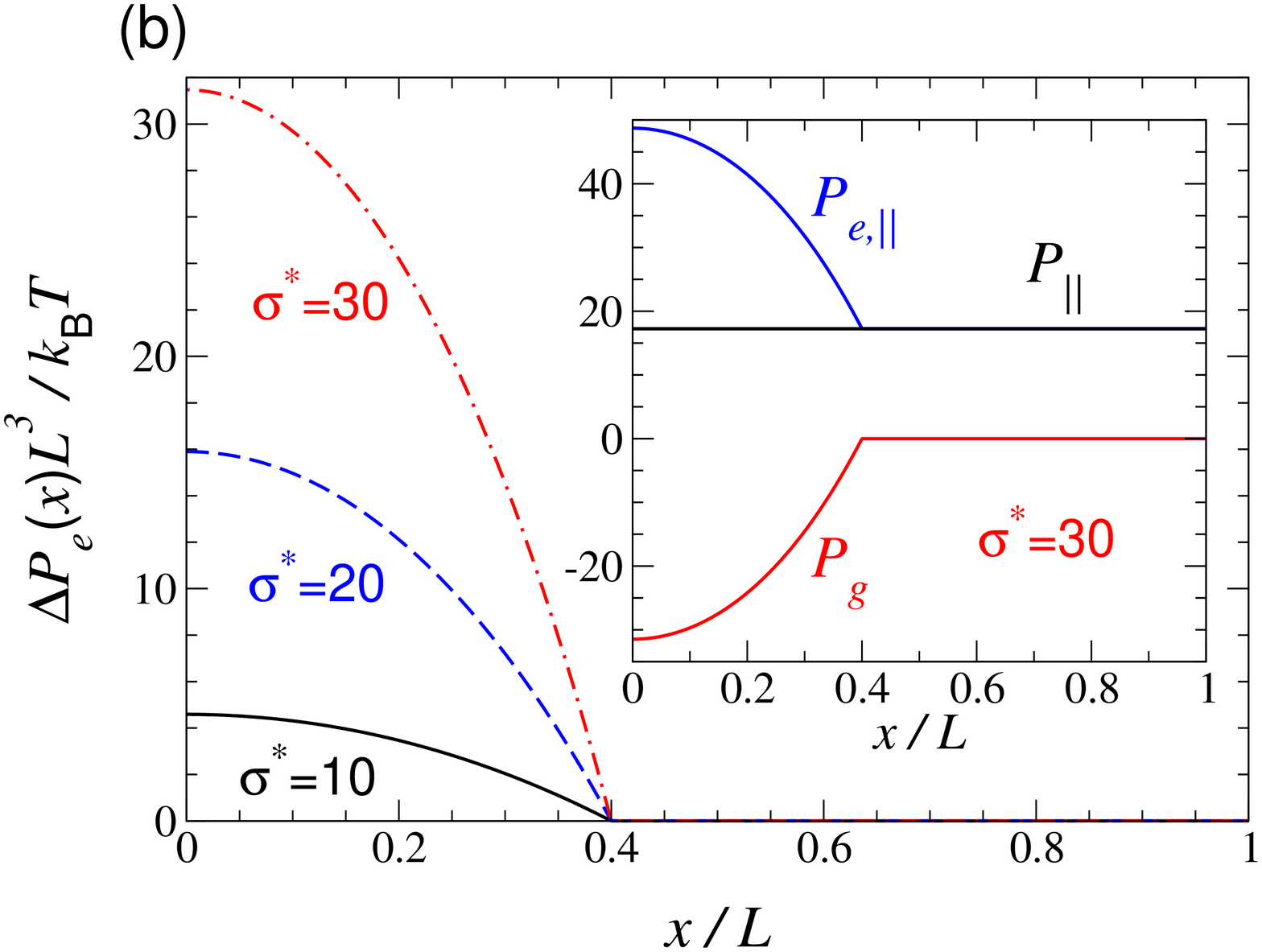}
\vspace*{-0.2cm}
\caption{
Electrostatic component of osmotic pressure $\Delta P_e(x)$ vs.~distance $x$ 
from left wall in the planar cell model for (a) surface-charged
[Eq.~(\ref{delta-Pe-flat-gel1}) or (\ref{delta-Pe-flat-gel2})],
(b) volume-charged [Eq.~(\ref{deltaPe-plane-gel})] planar microgels 
with system parameters of Fig.~\ref{psix}. Insets show profiles of normal component
of pressure $P_{\parallel}$ [Eq.~(\ref{P-plane2})] and electrostatic [Eq.~(\ref{Pe-plane1})]
and gel [Eq.~(\ref{Pg-plane})] contributions for reduced surface charge density $\sigma^*=30$.  
}\label{pex}
\end{figure}

\begin{figure}[t]
\includegraphics[width=\columnwidth]{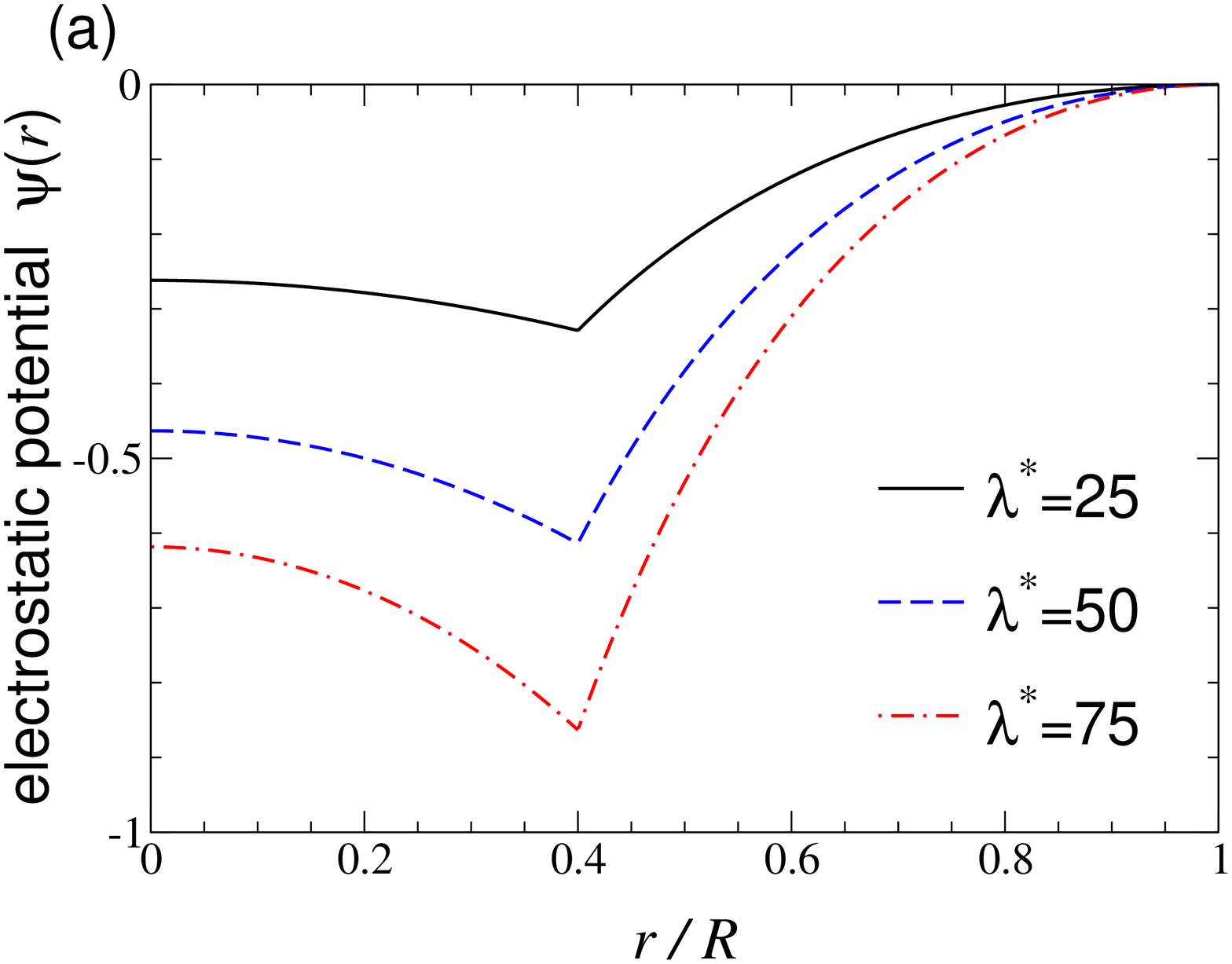}
\includegraphics[width=\columnwidth]{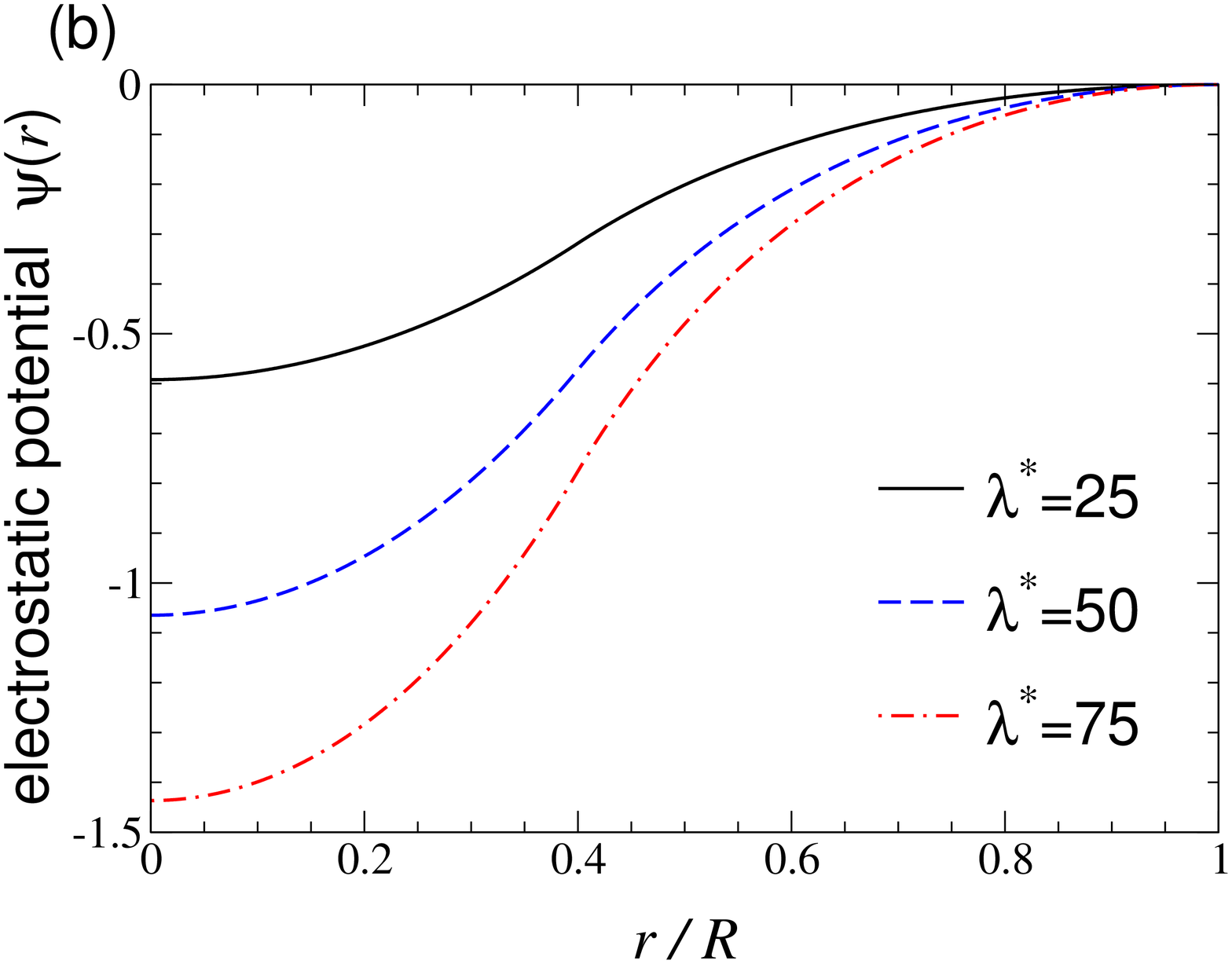}
\vspace*{-0.2cm}
\caption{
Reduced electrostatic potential $\psi(r)$ (shifted to 0 at $r=R$) vs.~radial distance $r$ from 
axis of a cylindrical cell of radius $R=50$ nm for (a) surface-charged, (b) volume-charged microgels 
of swollen radius $a=20$ nm and reduced valence per unit length $\lambda^*\equiv\lambda R=25$, 50, 75 
(solid, dashed, dash-dotted curves) 
at $T=293$ K in aqueous solution ($\lambda_B=0.714$ nm) with negligible salt.
}\label{psir-cylindrical}
\end{figure}

\begin{figure}[t]
\includegraphics[width=\columnwidth]{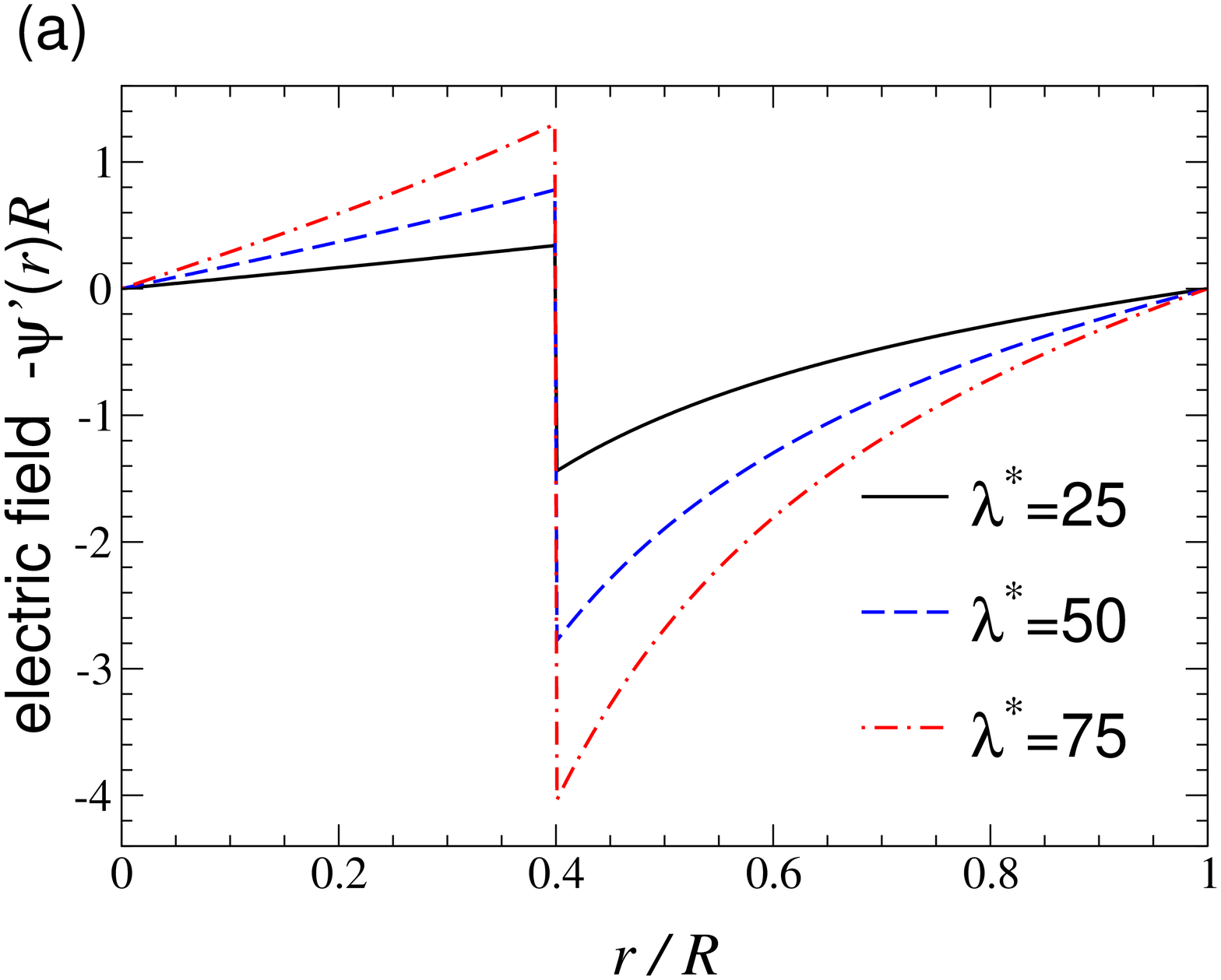}
\includegraphics[width=\columnwidth]{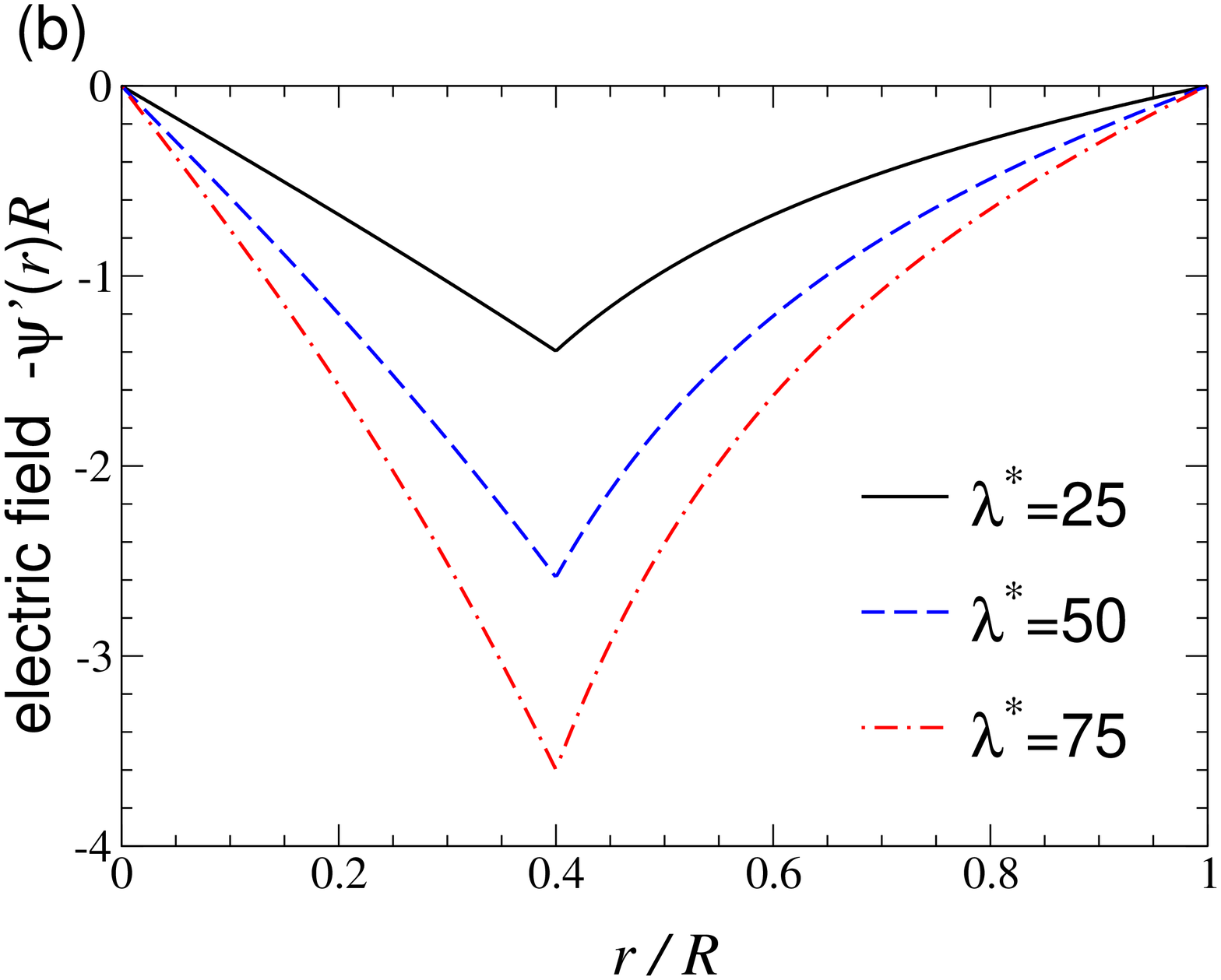}
\vspace*{-0.2cm}
\caption{
Reduced electric field $-\psi'(r)R$ vs.~radial distance $r$ from axis of a cylindrical cell for
(a) surface-charged, (b) volume-charged microgels with system parameters of Fig.~\ref{psir-cylindrical}.
}\label{Er-cylindrical}
\end{figure}

\begin{figure}[t]
\includegraphics[width=\columnwidth]{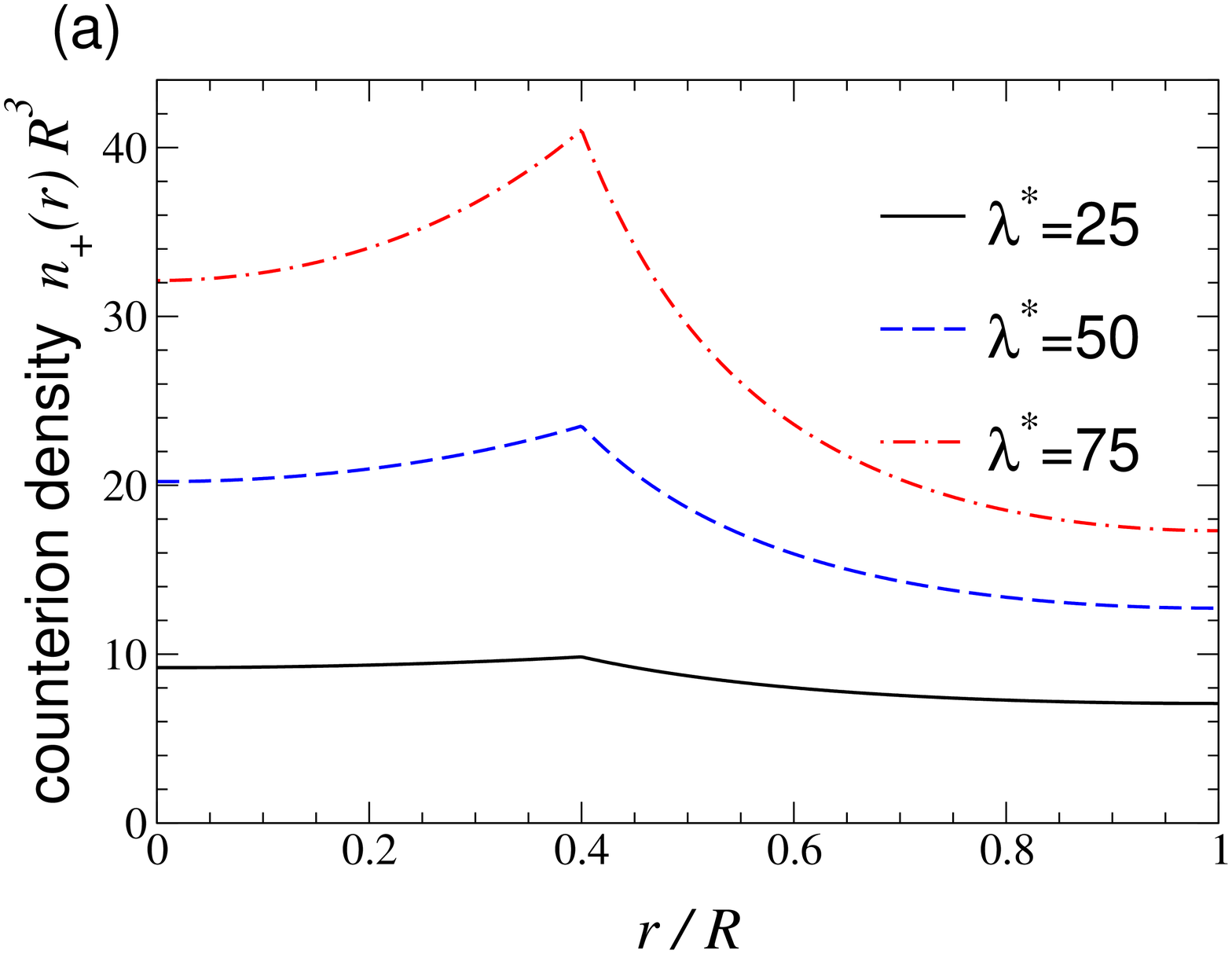}
\includegraphics[width=\columnwidth]{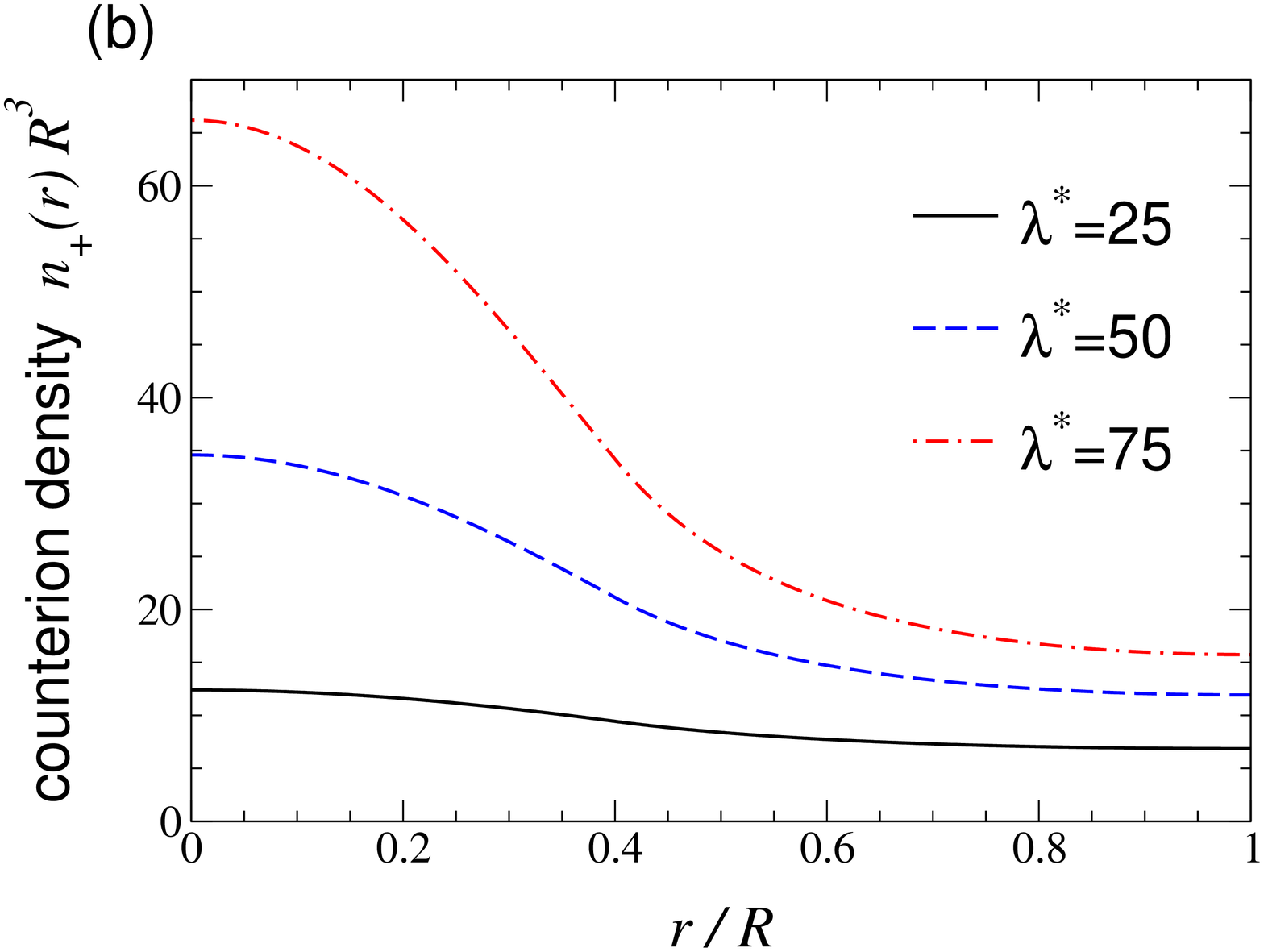}
\vspace*{-0.2cm}
\caption{
Counterion density $n_+(r)$ vs.~radial distance $r$ from axis of a cylindrical cell for
(a) surface-charged, (b) volume-charged microgels with system parameters of Fig.~\ref{psir-cylindrical}.
}\label{ncr-cylindrical}
\end{figure}

\begin{figure}[t]
\includegraphics[width=\columnwidth]{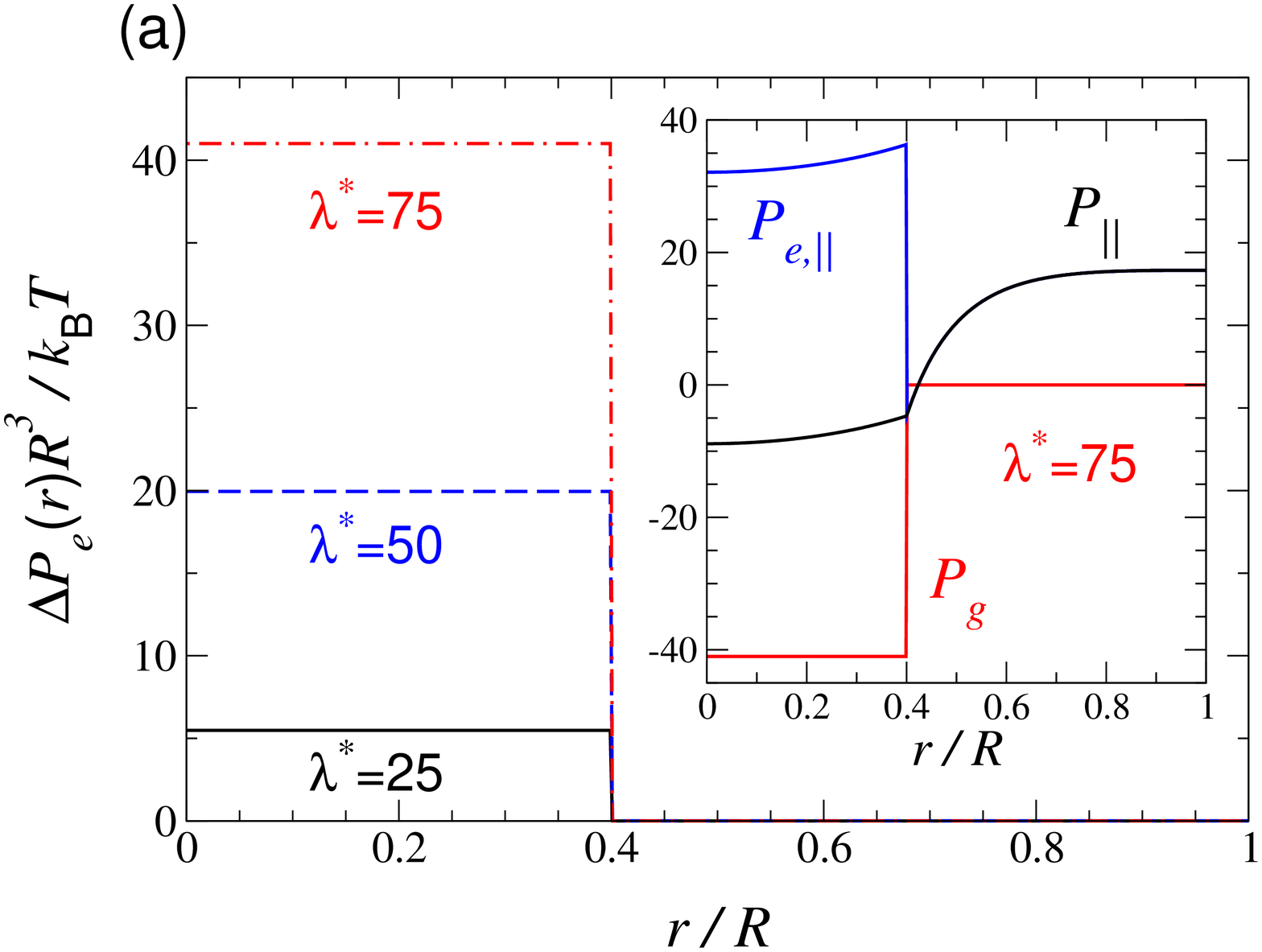}
\includegraphics[width=\columnwidth]{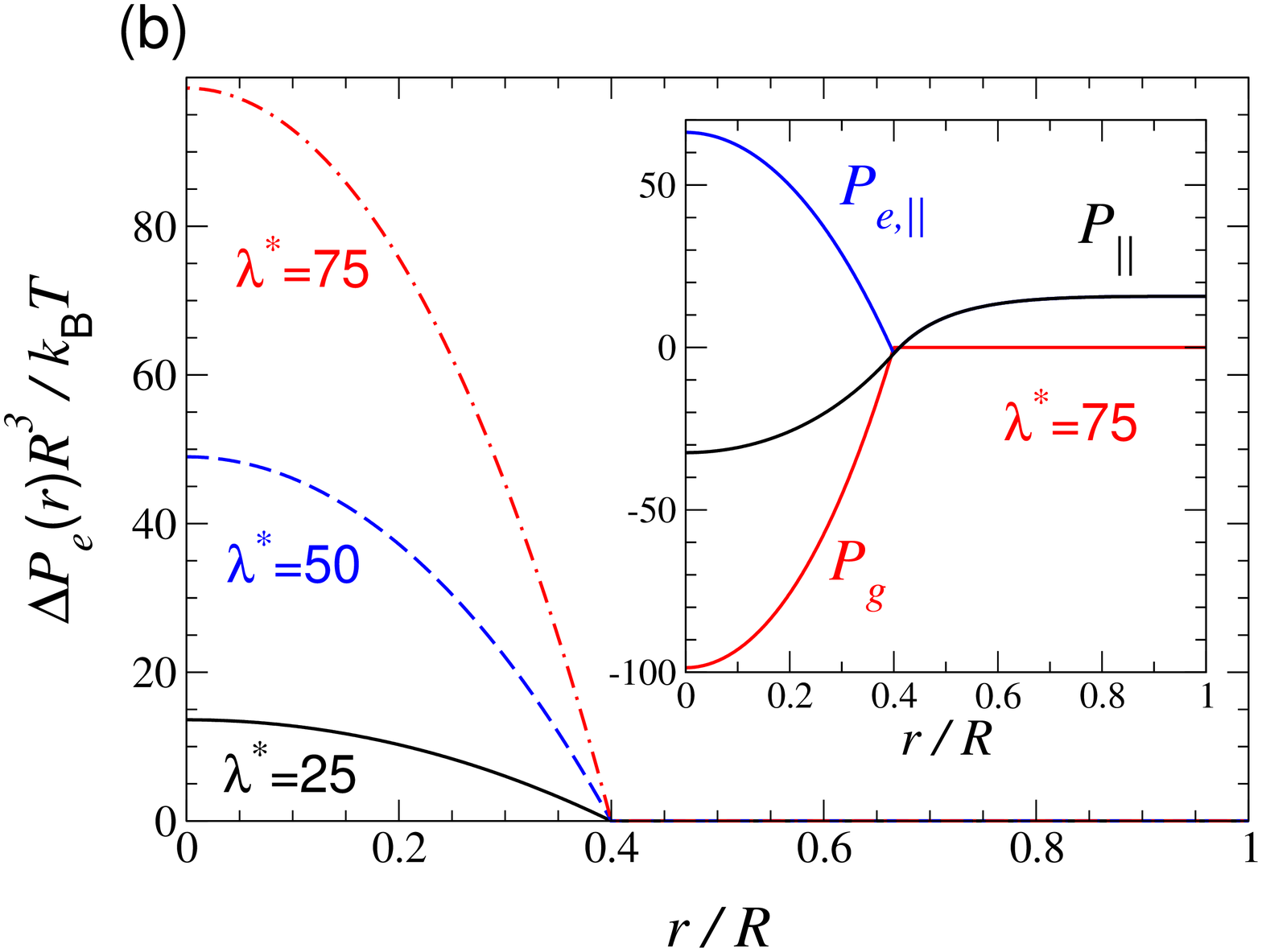}
\vspace*{-0.2cm}
\caption{
Electrostatic component of osmotic pressure $\Delta P_e(r)$ vs.~radial distance $r$ 
from axis of a cylindrical cell for (a) surface-charged
[Eq.~(\ref{delta-P-mechanical-cylinder-surface-charge})],
(b) volume-charged [Eq.~(\ref{deltaPe-cylinder-gel})] cylindrical microgels 
with system parameters of Fig.~\ref{psir-cylindrical}. Insets show radial profiles of 
normal component of total pressure $P_{\parallel}$ [Eq.~(\ref{P-normal-cylinder})]
and electrostatic [Eq.~(\ref{Pe-cylinder2a})] and gel [Eq.~(\ref{Pgr})] contributions 
for reduced linear charge per unit length $\lambda^*=75$.
}\label{per-cylindrical}
\end{figure}

\begin{figure}[t]
\includegraphics[width=\columnwidth]{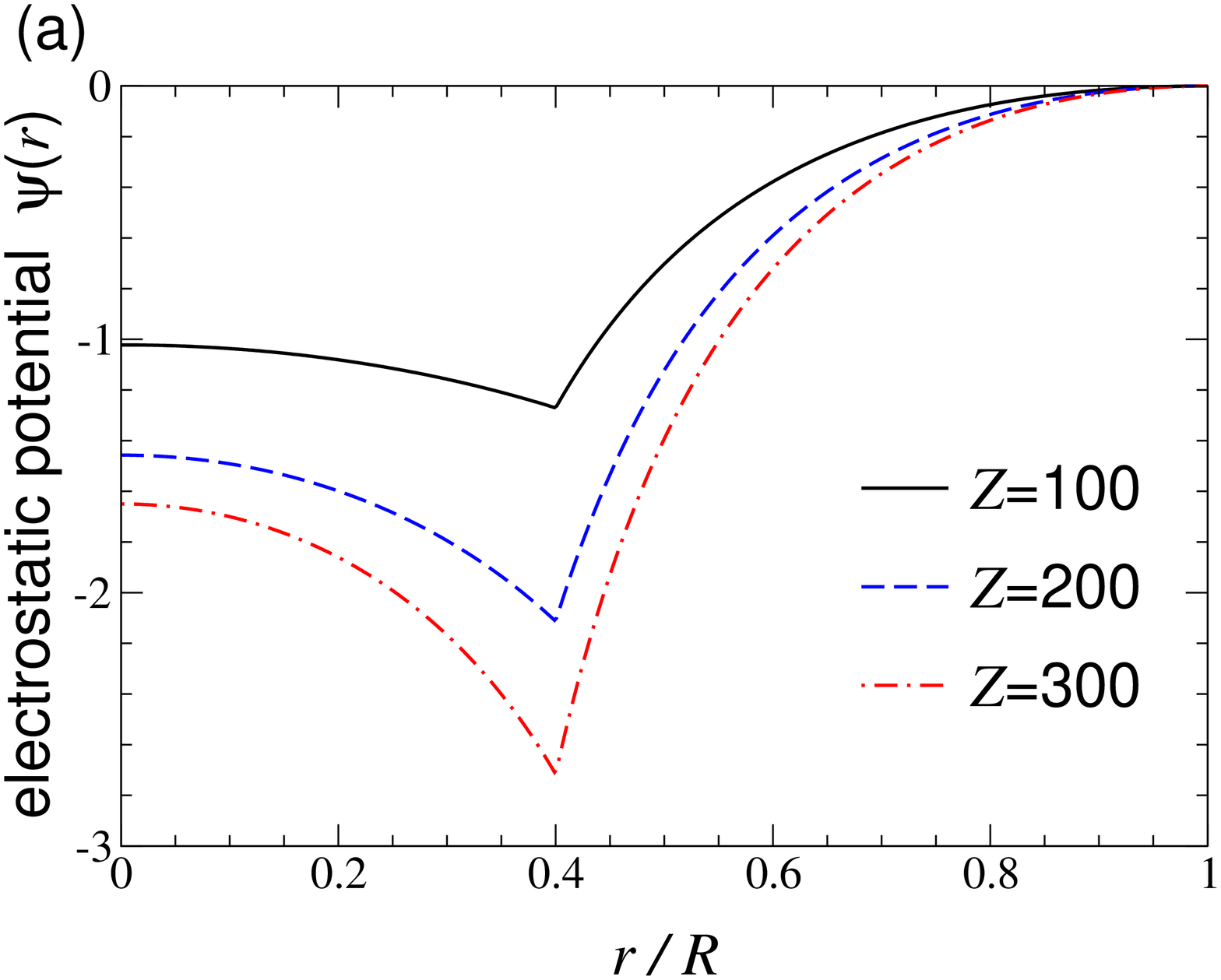}
\includegraphics[width=\columnwidth]{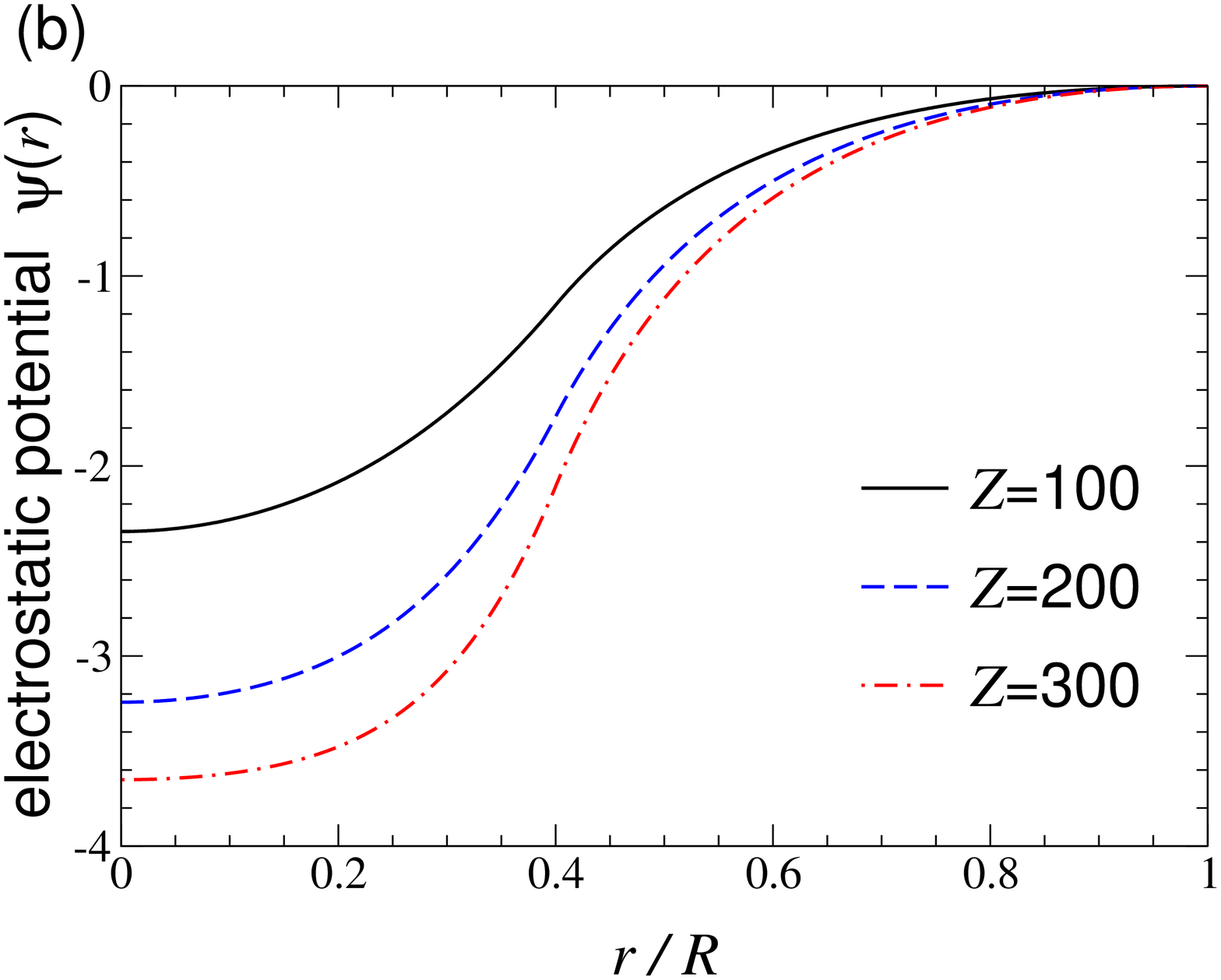}
\vspace*{-0.2cm}
\caption{
Reduced electrostatic potential $\psi(r)$ (shifted to 0 at $r=R$) vs.~radial distance $r$ from 
center of a spherical cell of radius $R=50$ nm for (a) surface-charged, (b) volume-charged microgels 
of swollen radius $a=20$ nm and valence $Z=100$, 200, 300 (solid, dashed, dash-dotted curves) 
at $T=293$ K in aqueous solution ($\lambda_B=0.714$ nm) with negligible salt.
}\label{psir}
\end{figure}

\begin{figure}[t]
\includegraphics[width=\columnwidth]{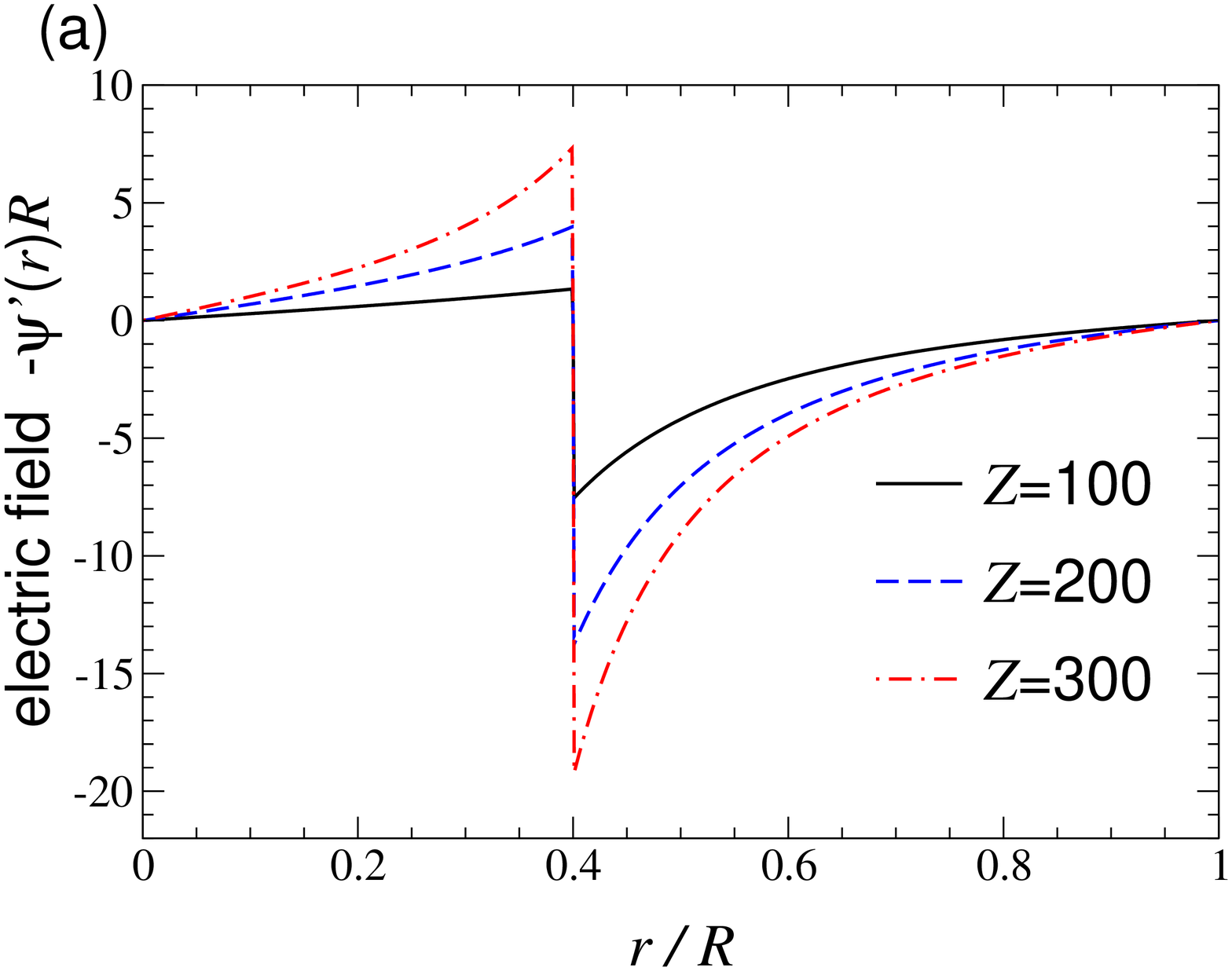}
\includegraphics[width=\columnwidth]{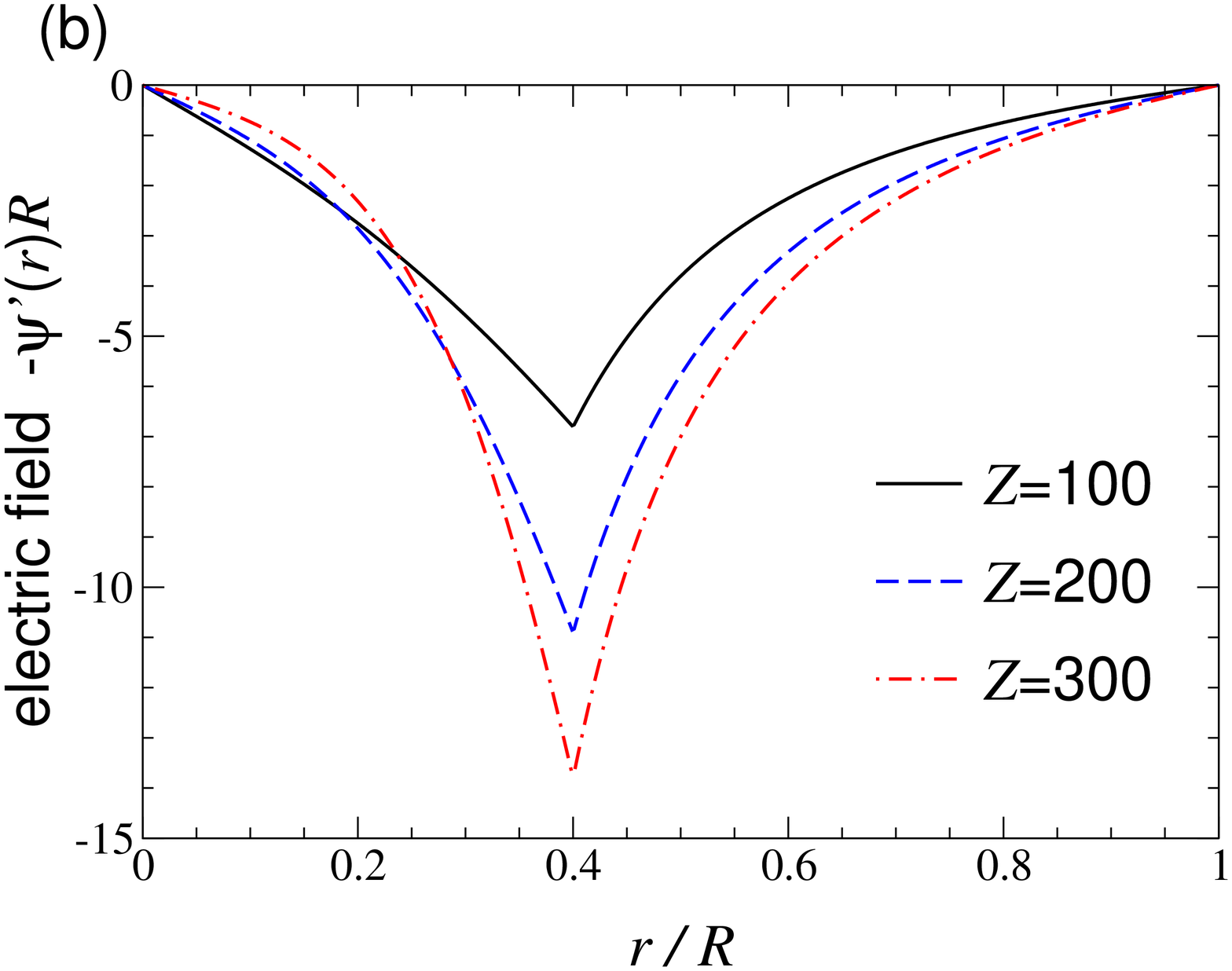}
\vspace*{-0.2cm}
\caption{
Reduced electric field $-\psi'(r)R$ vs.~radial distance $r$ from center of a spherical cell for
(a) surface-charged, (b) volume-charged microgels with system parameters of Fig.~\ref{psir}.
}\label{Er}
\end{figure}

\begin{figure}[t]
\includegraphics[width=\columnwidth]{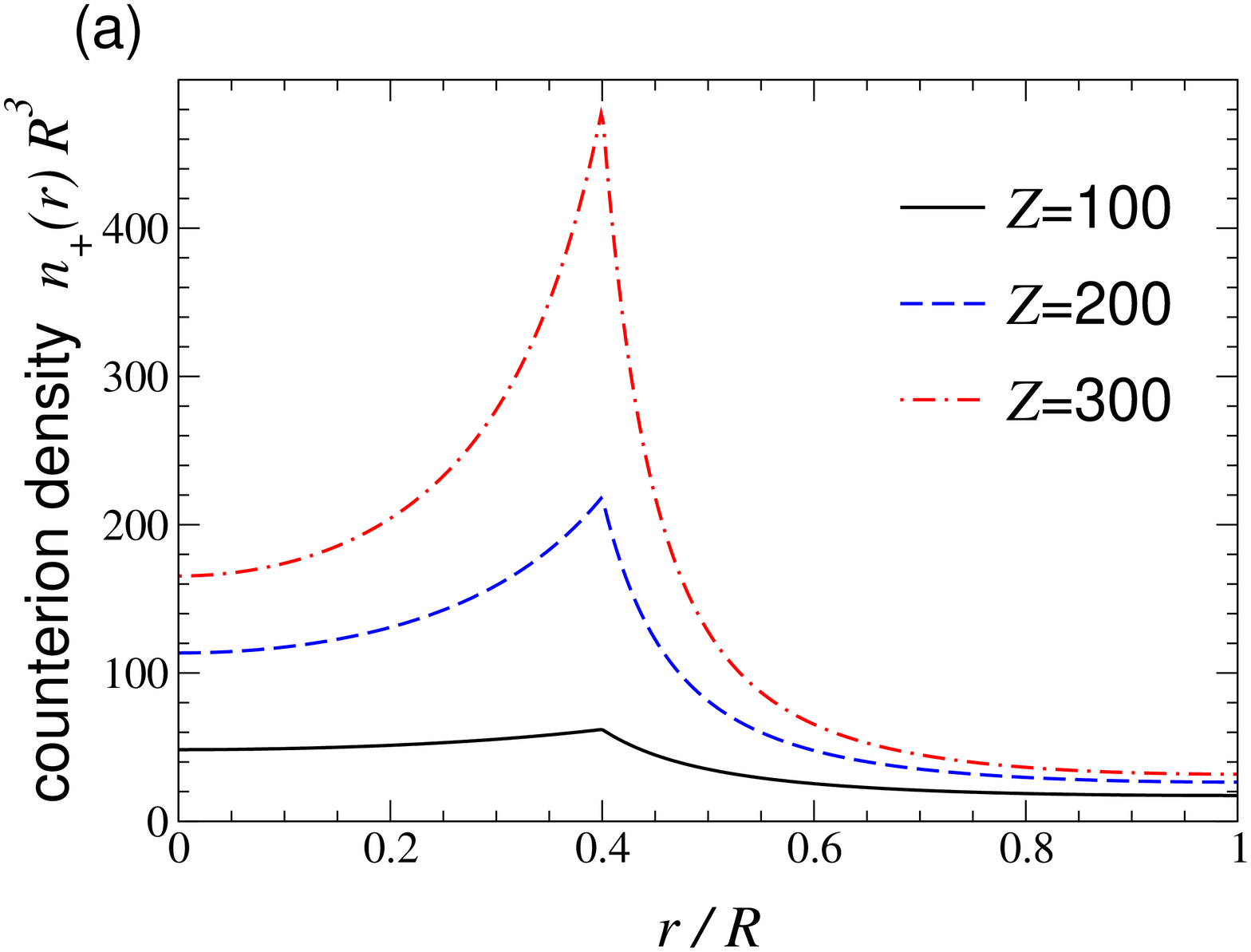}
\includegraphics[width=\columnwidth]{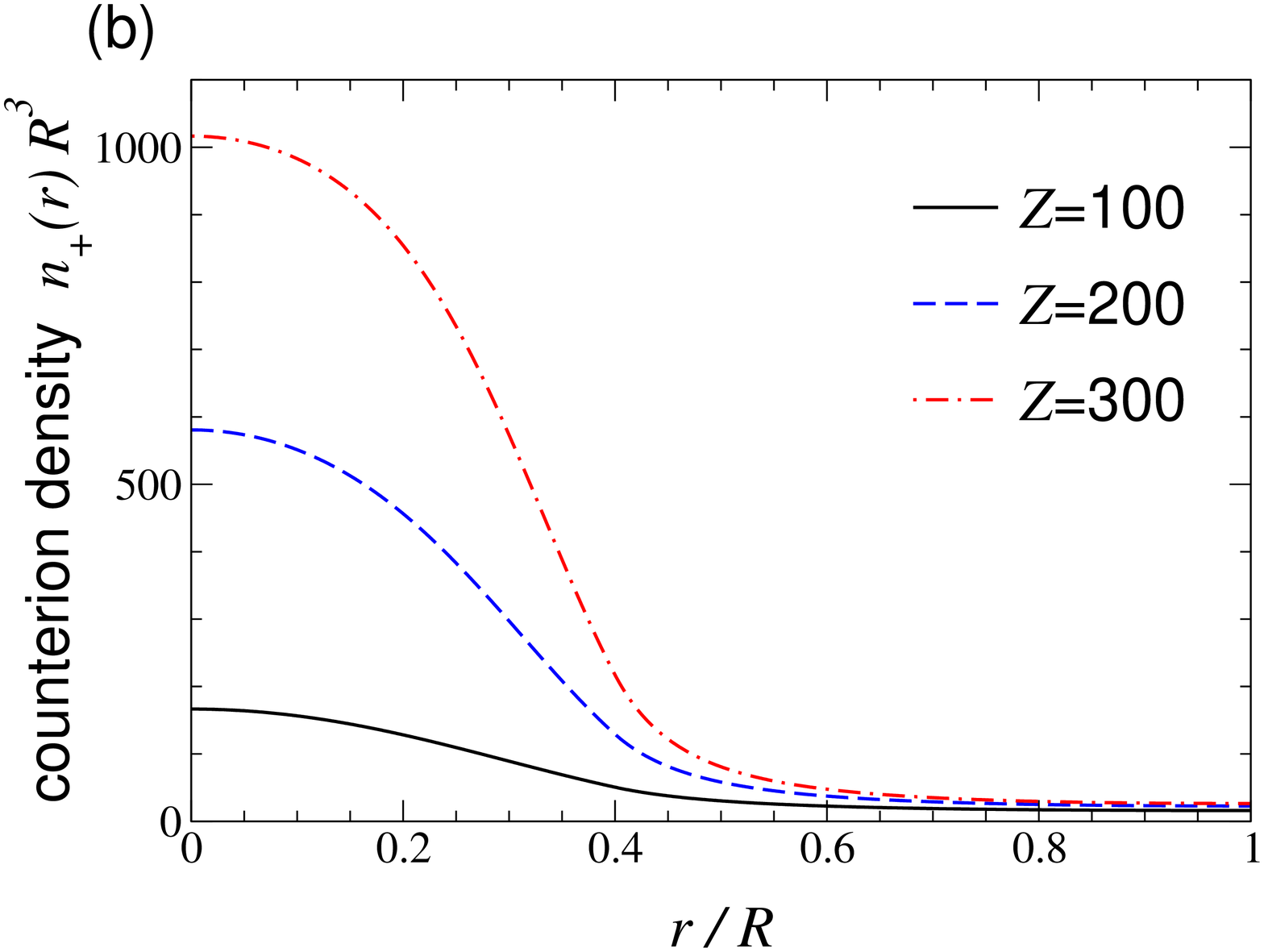}
\vspace*{-0.2cm}
\caption{
Counterion density $n_+(r)$ vs.~radial distance $r$ from center of a spherical cell for
(a) surface-charged, (b) volume-charged microgels with system parameters of Fig.~\ref{psir}.
}\label{ncr}
\end{figure}
%
\begin{figure}[t!]
\includegraphics[width=\columnwidth]{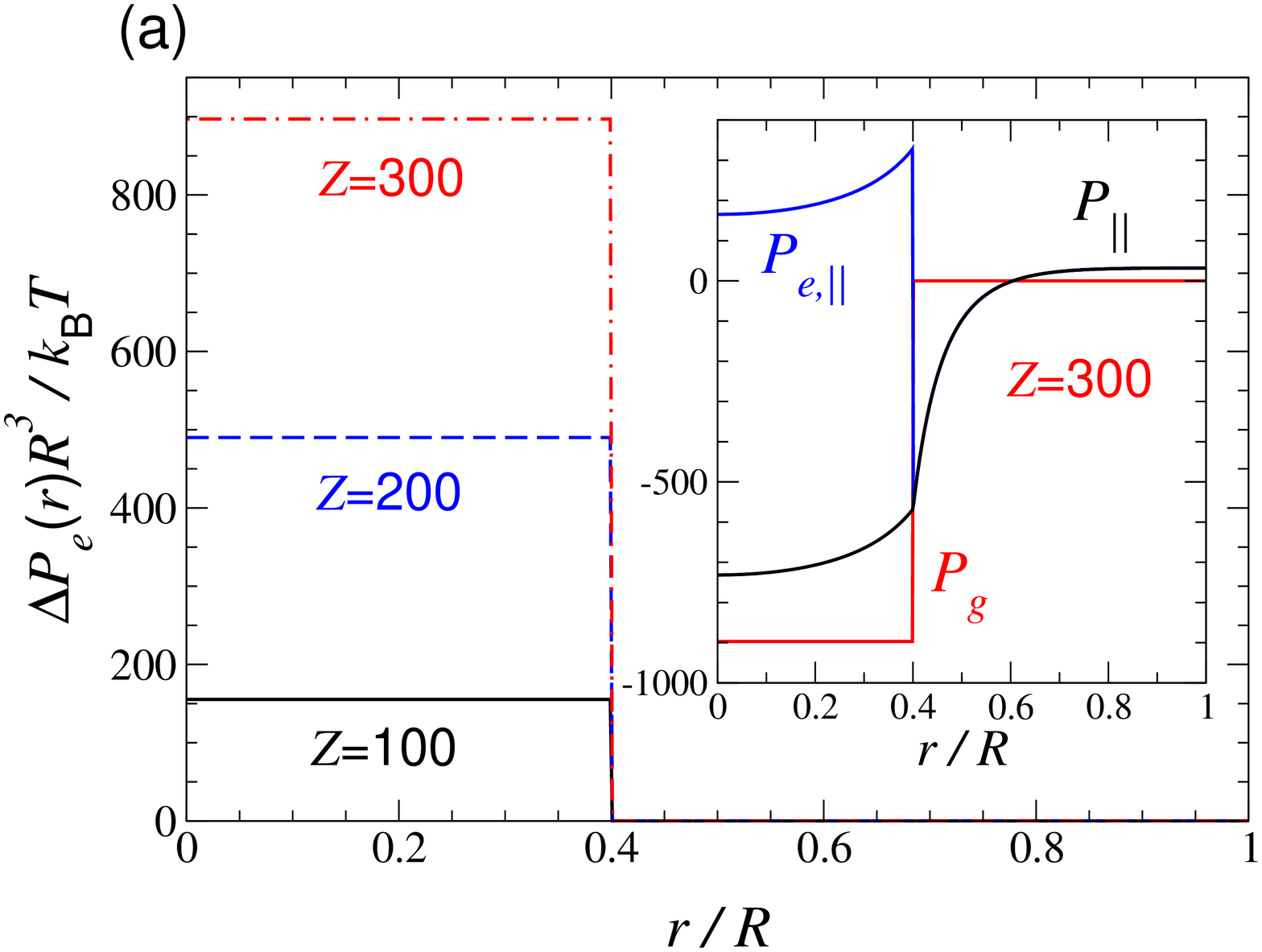}
\includegraphics[width=\columnwidth]{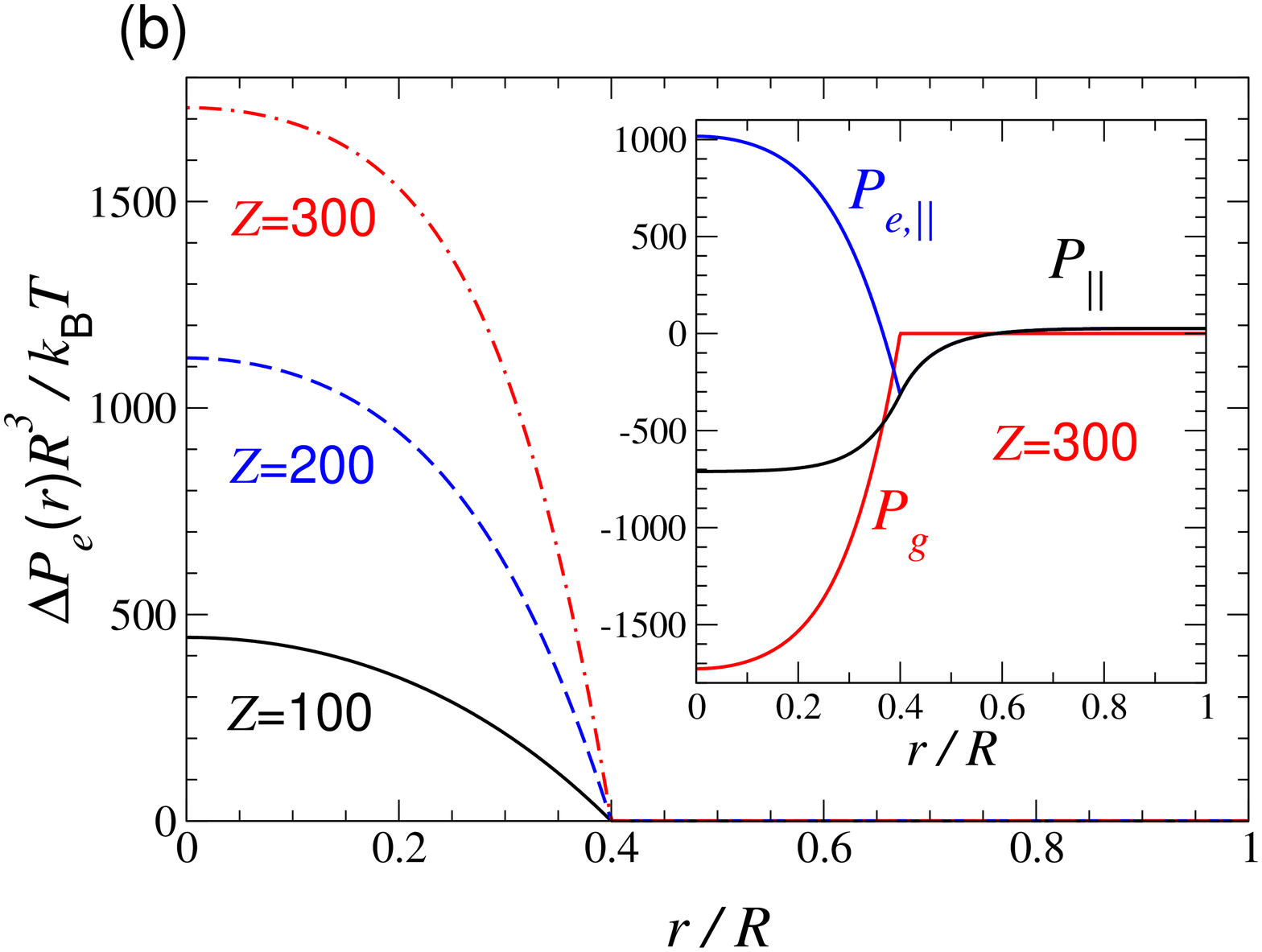}
\vspace*{-0.2cm}
\caption{
Electrostatic component of osmotic pressure $\Delta P_e(r)$ vs.~radial distance $r$ 
from center of a spherical cell for (a) surface-charged
[Eq.~(\ref{delta-P-mechanical-microgel-surface-charge})]
(b) volume-charged [Eq.~(\ref{deltaPe-sphere-gel})] spherical microgels 
with system parameters of Fig.~\ref{psir}. Insets show radial profiles of normal component
of total pressure $P_{\parallel}$ [Eq.~(\ref{Pr-sphere})] and electrostatic [Eq.~(\ref{Pe-sphere2a})]
and gel [Eq.~(\ref{Pgr})] contributions for valence $Z=300$.  
}\label{per-spherical}

\end{figure}

Typical numerical results for $\psi(x)$ [relative to $\psi(L)$], electric field $E(x)=-\psi'(x)$, 
counterion density $n_+(x)$, and electrostatic osmotic pressure $\Delta P_e(x)$ are shown in 
Figs.~\ref{psix}-\ref{pex}, respectively, for planar microgels at $T=293$ K in deionized
aqueous suspensions ($\lambda_B=0.714$ nm) with system parameters: 
swollen gel thickness $a=20$ nm, reduced valence/unit area $\sigma^*\equiv\sigma L^2=10$-30,
cell width $L=50$ nm, and reservoir salt concentration $c_r=10^{-7}$ M. 
For surface-charged microgels, the electrostatic potential and counterion density profiles 
exhibit cusps, and the electric field and electrostatic pressure are discontinuous, 
at the outer edge of the gel. [The electric field discontinuity is directly related to the 
electrostatic osmotic pressure via Eqs.~(\ref{delta-Pe-flat-gel1}) and (\ref{delta-Pe-flat-gel2}).]
In contrast, for volume-charged microgels, $\psi(x)$ and $n_+(x)$ are continuously differentiable, 
while $E(x)$ and $P_e(x)$ are continuous at the surface of the gel.
Note that the pressure (in $k_BT$ units) equals the counterion density at the edge of the cell.
With increasing $\sigma$, all of the profiles vary more rapidly with $x$ and the
electrostatic osmotic pressure of the microgel increases.

\subsection{Cylindrical and Spherical Ionic Microgels}\label{validation-microgels}

For cylindrical and spherical ionic microgels with axially or spherically symmetric
fixed charge distributions, we solved the PB equation in cylindrical ($d=2$) and
spherical ($d=3$) polar coordinates,
\begin{equation}
\psi''(r)+\frac{d-1}{r}\psi'(r)=\kappa^2\sinh\psi(r)+4\pi\lambda_B n_f(r),
\label{PB-polar}
\end{equation}
with boundary conditions $\psi'(0)=0$ (radial symmetry) and $\psi'(R)=0$ (electroneutrality).
Within Mathematica, we used the {\tt NDSolve} method to solve for the reduced 
electrostatic potential $\psi(r)$ in the interior ($r\le a$) and exterior ($r>a$) regions 
and the {\tt FindRoot} method to vary the electrostatic potential at the interface $\psi(a)$ 
to match the solutions in the two regions. 

For surface-charged cylindrical microgels, numerically evaluating 
Eq.~(\ref{delta-P-mechanical-cylinder-surface-charge}) yields precise agreement with 
the exact relation [Eq.~(\ref{cell-theorem-microgel-cylinder-surface-charge})].
Similarly, for surface-charged spherical microgels, evaluating 
Eq.~(\ref{delta-P-mechanical-microgel-surface-charge}) yields precise agreement with 
the corresponding exact relation [Eq.~(\ref{cell-theorem-microgel-surface-charge})].
Thus, our independently derived expressions for the electrostatic pressure inside
surface-charged microgels agree precisely with each other, but not with the 
prediction of ref.~\cite{gasser2019} for spherical microgels. 

In passing, we note that our numerical solutions of the PB equation for impenetrable cylindrical 
and spherical charged colloids with surface boundary condition $\psi'(a)=4\pi\lambda_B\sigma$
confirm that Eq.~(\ref{P-hard-cylinder}) for cylinders and Eq.~(\ref{P-colloid}) for spheres 
yield bulk pressures obeying the cell theorem [Eq.~(\ref{cell-theorem-cylinder})] and
the derived integral relations [Eq.~(\ref{P-cylinder2b}) for cylinders, Eq.~(\ref{P-sphere2b}) 
for spheres].

For volume-charged microgels, our numerical solution of Eq.~(\ref{PB-polar}) confirms that the 
electrostatic osmotic pressure is {\it nonuniform} inside the microgel, increasing continuously 
from the surface inward. Although the value at the cell center, $\Delta P_e(0)$, exceeds the 
electrostatic component of the exact electrostatic osmotic pressure 
[Eq.~(\ref{cell-theorem-microgel-cylinder-volume-charge}) or (\ref{cell-theorem-microgel-volume-charge})], 
quite remarkably, the {\it average} of the PB solution $\Delta P_e(r)$ over the microgel volume
precisely agrees with the exact relation. For a cylindrical microgel, the volume average of
Eq.~(\ref{deltaPe-cylinder-gel}),
\begin{equation}
\la \Delta P_e\ra=\frac{2\lambda}{\pi a^4}\int_0^a dr\, r[\psi(a)-\psi(r)],
\label{Delta-Pe-simple-cylinder-volume-charge}
\end{equation}
precisely agrees with Eq.~(\ref{cell-theorem-microgel-cylinder-volume-charge}),
while for a spherical microgel, the volume average of Eq.~(\ref{deltaPe-sphere-gel}),
\begin{equation}
\la \Delta P_e\ra=\frac{9Z}{4\pi a^6}\int_0^a dr\, r^2[\psi(a)-\psi(r)],
\label{Delta-Pe-simple-volume-charge}
\end{equation}
precisely agrees with Eq.~(\ref{cell-theorem-microgel-volume-charge}). Thus, we conclude that,
quite generally for all geometries, the exact result for the electrostatic component of the 
osmotic pressure of an ionic microgel with uniformly distributed volume charge equals the 
volume average of the electrostatic osmotic pressure profile predicted by PB theory in the cell model.

Illustrative numerical results for the reduced electrostatic potential $\psi(r)$ 
[relative to $\psi(R)$], the reduced electric field $E(r)=-\psi'(r)$, the counterion 
number density $n_+(r)$, and the electrostatic osmotic pressure $\Delta P_e(r)$ 
are shown in Figs.~\ref{psir-cylindrical}-\ref{per-cylindrical} for 
cylindrical microgels with reduced valence per unit length $\lambda^*\equiv\lambda R=25$-75
and in Figs.~\ref{psir}-\ref{per-spherical} for spherical microgels with valence $Z=100$-300
with swollen radius $a=20$ nm in a cell of radius $R=50$ nm at $T=293$ K in deionized 
aqueous suspensions ($\lambda_B=0.714$ nm) and salt reservoir concentration $c_r=10^{-7}$ M. 
Also shown in Figs.~\ref{per-cylindrical} and \ref{per-spherical} (insets) are 
the normal component of the pressure and its electrostatic and gel contributions,
confirming radial variation of $P_{\parallel}(r)$ and $P_{e,\parallel}(r)$.

The results are qualitatively similar to those obtained in Sec.~\ref{validation-planar-gels} 
for planar microgels. For surface-charged microgels, $\psi(r)$ and $n_+(r)$ exhibit cusps at the 
microgel surface, consistent with the Monte Carlo simulation data of Scotti \etalia\cite{nieves-pnas2016}
for the same model of spherical microgels, while $E(r)$ and $P_{e,\parallel}(r)$ are 
discontinuous at $r=a$. The electric field discontinuity is directly related to the 
electrostatic osmotic pressure via Eq.~(\ref{delta-P-mechanical-cylinder-surface-charge}) 
for cylinders and Eq.~(\ref{delta-P-mechanical-microgel-surface-charge}) for spheres.
In contrast, for volume-charged microgels, $\psi(r)$ and $n_+(r)$ are continuously differentiable, 
consistent with the molecular dynamics simulation data of Claudio \etalia\cite{holm2009} and of 
Denton and Tang~\cite{denton-tang2016} for the same model of spherical microgels, while $E(r)$ and 
$P_{e,\parallel}(r)$ are continuous at the surface of the microgel. With increasing $\lambda$ or $Z$, 
the profiles all vary more rapidly with $r$ and the electrostatic osmotic pressure increases.

A notable difference between the results in planar and curved geometries is that the 
normal component of the total pressure $P_{\parallel}$ is spatially uniform in the planar cell,
but nonuniform in the cylindrical and spherical cells. In curved geometries, $P_{\parallel}(r)$
can even become negative, as long as $\nabla\cdot{\bf P}=0$ [Eqs.~(\ref{divergence-cylinder})
or (\ref{divergence-sphere})]. Of course, the transverse component of the electrostatic
pressure $P_{e,\bot}(r)$ [Eq.~(\ref{Pe-transverse})] is strictly positive.
Note that, while the normal component of the electrostatic pressure (in $k_BT$ units) equals 
the microion density at the edge and center of the cell, where the electric field vanishes,
the electrostatic osmotic pressure is not equal to the pressure difference, 
$|P_{e,\parallel}(0)-P_{e,\parallel}(R)|$, between the center and edge of the cell, 
but rather depends on the variation of $P_{e,\parallel}(r)$ relative to $P_{\parallel}(r)$ 
inside the microgel. Thus, our prediction for $\Delta P_e$ differs from that of 
ref.~\cite{gasser2019}.

The validity of the results derived in Secs.~\ref{applications} and \ref{exact} for the 
electrostatic component of the osmotic pressure of ionic microgels is further confirmed 
by molecular dynamics simulations of the primitive model previously performed in a 
spherical cell with explicit monovalent counterions for volume-charged spherical 
microgels~\cite{denton-tang2016}.
Counterion density profiles extracted from these simulations and from the solution
of the PB equation, when used to explicitly evaluate the statistical mechanical relation 
for the electrostatic osmotic pressure [Eq.~(\ref{cell-theorem-microgel-volume-charge})],
yielded excellent agreement, thus demonstrating the accuracy of PB theory 
in the spherical cell model for systems with monovalent counterions and no salt.

\begin{figure}[t!]
\includegraphics[width=\columnwidth]{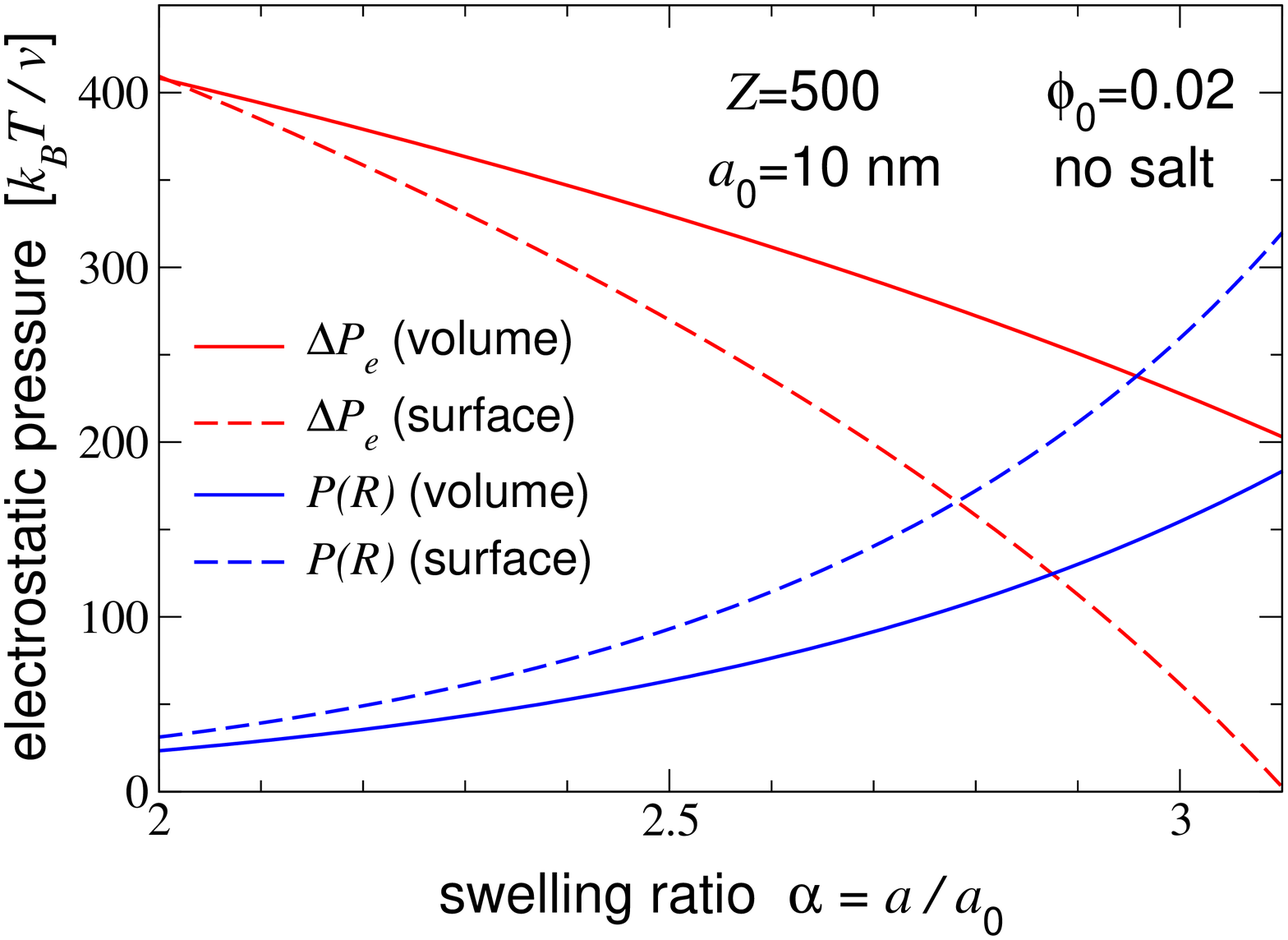}
\vspace*{-0.2cm}
\caption{
Exact electrostatic osmotic pressure of an ionic microgel $\Delta P_e$ (red curves) 
and bulk pressure of suspension $P(R)$ (blue curves), in units of $k_BT$/swollen particle volume 
($v=4\pi a^3/3$), vs.~radial swelling ratio $\alpha$ for surface-charged (dashed curves) and
volume-charged (solid curves) microgels with valence $Z=500$, collapsed radius $a_0=10$ nm,
and collapsed volume fraction $\phi_0=0.02$ in the spherical cell model of a salt-free suspension.
}\label{pe-vs-alpha-microgel}
\end{figure}

To demonstrate the relevance of our analysis for swelling of spherical ionic microgels, 
we present results for the electrostatic component of the osmotic pressure $\Delta P_e$ 
vs.~radial swelling ratio $\alpha$.
Figure~\ref{pe-vs-alpha-microgel} shows sample results for surface- and volume-charged microgels 
of valence $Z=500$, collapsed radius $a_0=10$ nm, and collapsed volume fraction $\phi_0=(a_0/R)^3=0.02$.
Also shown is the bulk pressure of the suspension $P(R)$. With increasing $\alpha$ 
(i.e., increasing volume fraction), $\Delta P_e$ monotonically decreases, while $P(R)$ 
steadily increases.  When $\alpha$ reaches a value at which the normal components of the 
electrostatic pressure and the gel pressure [e.g., from Eq.~(\ref{Flory-p})] are equal 
in magnitude, the total osmotic pressure vanishes (i.e., $P_{\parallel}(r)$ is continuous) 
and the microgel is in equilibrium~\cite{denton-tang2016}. 
The stable swollen size depends on the distribution of fixed charge.

A microgel with fixed charge uniformly distributed over its volume has an electrostatic component 
of the osmotic pressure that is consistently higher than that of a microgel with the 
same charge uniformly spread over its surface. The difference is largely due to a higher 
self-energy contribution to the electrostatic pressure for volume-charged microgels.
Thus, a volume-charged microgel attains an equilibrium swollen radius larger than that of a 
surface-charged microgel with the same collapsed radius and fixed charge. Simply put, 
volume-charged microgels tend to swell more than surface-charged microgels.
Also clear from Fig.~\ref{pe-vs-alpha-microgel} is that the pressure of a suspension of 
surface-charged microgels systematically exceeds that of a suspension of equally swollen 
volume-charged microgels. 

Spatial uniformity of the gel pressure inside a surface-charged microgel implies
uniform stretching of the polymer network and thus a uniform radial swelling ratio.
For volume-charged microgels, in contrast, $P_g(r)$ varies with radial distance, 
and thus the equilibrium radial swelling ratio should be nonuniform. However, nonuniform 
swelling would imply a spatially varying fixed charge density, which in turn would 
modify the electrostatic pressure profile. In previous work on swelling of volume-charged 
microgels~\cite{denton-tang2016}, we simply assumed $\alpha$ to be constant. A more refined 
analysis would self-consistently account for spatial variation of $P_g(r)$ by allowing
$\alpha$ to vary with $r$ and iterating between the electrostatic and gel pressure profiles 
until convergence. 

\begin{figure}[t]
\includegraphics[width=\columnwidth]{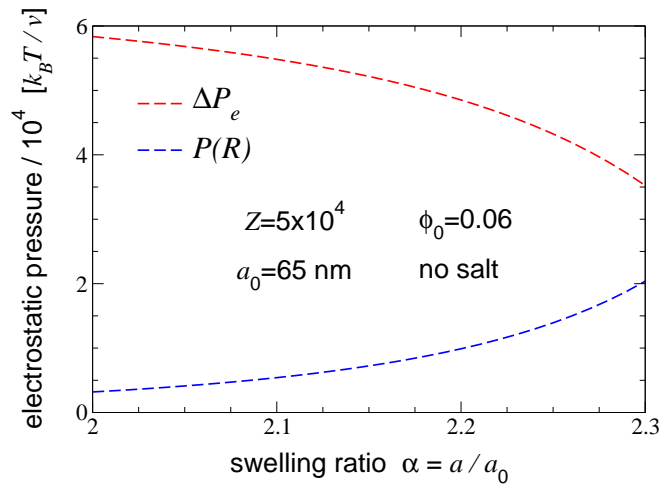}
\vspace*{-0.2cm}
\caption{
Exact electrostatic osmotic pressure of an ionic microgel $\Delta P_e$ (red curve) 
and bulk pressure of suspension $P(R)$ (blue curve), in units of $k_BT/v$, 
vs.~radial swelling ratio $\alpha$ for surface-charged microgels with valence $Z=5\times 10^4$, 
collapsed radius $a_0=65$ nm, 
and collapsed volume fraction $\phi_0=0.06$ in the spherical cell model of a salt-free suspension.
System parameters are comparable to those of ref.~\cite{nieves-pnas2016}.
}\label{pe-vs-alpha-expt}
\end{figure}

Finally, we attempt to achieve closer contact with experimental measurements of osmotic pressure and 
swelling of microgels. Figure~\ref{pe-vs-alpha-expt} shows our predictions for $\Delta P_e$ and 
$P(R)$ vs.~$\alpha$ for a suspension of surface-charged spherical microgels with system parameters 
roughly comparable to those in a recent experimental study of pNIPAM microgels~\cite{nieves-pnas2016}:
valence $Z=5\times 10^4$, collapsed radius $a_0=65$ nm, and collapsed volume fraction $\phi_0=0.06$. 
(As $\alpha$ varies from 2-2.3, the swollen volume fraction $\phi$ ranges from 0.48-0.73,
remaining below close packing.) 
Scotti~\etalia\cite{nieves-pnas2016} estimated $Z=6.8\times 10^4$ based on mass balance during 
the synthesis, although this value is likely an upper limit, as it assumes complete dissociation 
of the initiator, which is believed to be confined to the particle periphery. Over the range of 
$\alpha$ covered in Fig.~\ref{pe-vs-alpha-expt}, $\Delta P_e$ falls from roughly 25 to 10 kPa, 
while $P(R)$ rises from about 1 to 6 kPa, a range comparable to the pressures measured by osmometry 
for this system~\cite{nieves-pnas2016}.
At concentrations beyond close packing, steric interparticle interactions should 
increase the pressure of the suspension~\cite{urich-denton2016,weyer-denton2018}.

\section{Conclusions}\label{conclusions} 

In summary, we used two complementary methods to derive fundamental relations, valid within 
the cell model, for the electrostatic contribution to the osmotic pressure of penetrable macroions,
which are permeable to solvent and mobile microions, with a given distribution of fixed charge.
In Poisson-Boltzmann theory, we obtained the bulk pressure of a suspension by minimizing 
the semi-grand potential functional with respect to the electrostatic potential. 
In mechanical equilibrium, the electrostatic contribution to the normal component of the 
pressure in the cell, obtained from the electrostatic pressure tensor, must be balanced by a 
counteracting pressure acting on the fixed charge, maintaining continuity of the total pressure.
Our approach essentially disentangles the electrostatic and gel contributions to the total pressure.
We applied the expressions derived from PB theory to idealized models of ionic microgels.
In a statistical mechanical approach, based on the Hamiltonian and partition function,
we varied the semi-grand potential with respect to the thickness of a microgel slab or the 
radius of a cylinder or sphere to obtain exact relations for the microgel electrostatic 
osmotic pressure.

We presented explicit expressions for the electrostatic osmotic pressure of planar membranes 
and of planar, cylindrical, and spherical microgels with fixed charge uniformly spread over 
the particle surface or volume. These expressions may be regarded as analogues for 
penetrable macroions of the contact value theorem for planar charged surfaces.
In equilibrium, the normal component of the electrostatic pressure $P_{e,\parallel}(r)$ of a 
surface-charged microgel is discontinuous at the surface, while for a volume-charged microgel, 
$P_{e,\parallel}(r)$ is continuous, increasing from the surface inward. In all cases, however,
the gel pressure acting on the fixed charge compensates $P_{e,\parallel}(r)$ such that the 
normal component of the total pressure $P_{\parallel}(r)$ remains continuous. Our analysis 
demonstrates that the electrostatic osmotic pressure of a cylindrical or spherical microgel 
is not determined by the difference in microion density between the center and edge of the cell,
but instead depends on the variation of $P_{e,\parallel}(r)$ relative to $P_{\parallel}(r)$ 
inside the microgel.

To validate our results, we numerically solved the PB equation in the cell model with planar, 
cylindrical, and spherical symmetries. We confirmed that, for surface-charged microgels, 
the two approaches yield identical pressures, while for volume-charged microgels, the exact 
osmotic pressure precisely equals the volume average of the spatially varying PB pressure.
To illustrate the relevance of our derivations for swelling of ionic microgels, we computed 
counterion densities and electrostatic pressures for typical experimental system parameters.
Predicting the equilibrium swelling ratio of an ionic microgel with a given distribution of 
fixed charge requires adding the electrostatic pressure to the gel pressure of the swollen 
microgel, which may be independently obtained from simulations or theory. Future applications 
to swelling of microgels and other soft, permeable colloids may combine our approach for 
modeling the electrostatic osmotic pressure with theoretical and computational modeling of 
the gel osmotic pressure in
crosslinked polymer networks~\cite{winkler2014,winkler2017,zaccarelli2017} and 
polyelectrolyte gels~\cite{linse2002,moncho-jorda2013,quesada-perez2013,potemkin2015,
moncho-jorda-dzubiella2016,schneider2017,holm2017,quesada-perez-moncho-jorda2018,alexeev2018}.

\acknowledgments
This work was partially supported by the National Science Foundation (Grant No.~DMR-1106331).
For helpful discussions, we acknowledge Sylvio May, Gerhard N\"agele, Jan Dhont, Mariano Brito, 
Alberto Fern\'andez-Nieves, Urs Gasser, and Andrea Scotti.
MOA thanks Shaqra University for financial support.



\end{document}